\documentclass[%
 reprint,
nofootinbib,
 amsmath,amssymb,
 aps,
 prd,
 floatfix
]{revtex4-2}
\usepackage{ragged2e}
\usepackage{subcaption}   
\DeclareCaptionJustification{justified}{\justifying}
\usepackage{color}

\usepackage{graphicx}% Include figure files
\usepackage{dcolumn}% Align table columns on decimal point
\usepackage{bm}% bold math
\usepackage{hyperref}
        
\usepackage{orcidlink}

\def\diff{\mathrm{d}}

\DeclareMathOperator{\const}{const}

\def\1{\mathbb{1}} % Bold 1
\def\R{\mathbb{R}}

\def\am{\mathrm{am}} %JacobiAmplitude
\def\sn{\mathrm{sn}} %JacobiSN
\def\cn{\mathrm{cn}} %JacobiCN
\def\dn{\mathrm{dn}} %JacobiDN
\newcommand\restr[2]{  \left.\kern-\nulldelimiterspace  #1  \vphantom{\big|}  \right|_{#2} } % restriction sign

\usepackage{mathtools}
\mathtoolsset{showonlyrefs} % requires nofootinbib class option

\begin{document}

\title{Looking through the Kerr disk}

\author{Maciej Maliborski\hspace{0.233ex}\orcidlink{0000-0002-8621-9761}}
\affiliation{University of Vienna, Faculty of Mathematics, Oskar-Morgenstern-Platz 1, 1090 Vienna, Austria}
\affiliation{University of Vienna, Gravitational Physics, Boltzmanngasse 5, 1090 Vienna, Austria}
\affiliation{TU Wien, Institute of Analysis and Scientific Computing, Wiedner Hauptstraße 8-10, 1040 Vienna, Austria}

\author{Tobias C. Sutter\hspace{0.233ex}\orcidlink{0009-0008-3179-7979}}
\email{tobias.christoph.sutter@univie.ac.at}
 \affiliation{ University of Vienna, Faculty of Physics, Währingerstrasse 17, 1090 Vienna.
}

\date{\today}

\begin{abstract}
We study null geodesics that connect the two asymptotically flat regions of the maximally extended Kerr spacetime. These vortical geodesics traverse both horizons and pass through the ring singularity, linking the positive-$r$ exterior to the negative-$r$ asymptotic side. Using impact parameters, we identify a closed subset of parameter space, the inner throat, where the radial potential has no real roots, and photons exhibit no radial turning points. In this region, at most two constant-latitude geodesics exist, one of which is aligned with the principal null direction. We also identify the forbidden polar-angle band that limits the range of geodesics reaching an asymptotic observer.
We solve the geodesic equations analytically and numerically in Eddington-Finkelstein-like coordinates, obtaining mutually consistent results that correct and extend previously available formulae.
The resulting trajectories are used to construct simulated views for an observer in the negative-$r$ domain, revealing strong image distortion and inversion, with possible implications for analogous white-hole configurations.
\end{abstract}

\maketitle

%%%%%%%%%%%%%%%%%%%%%%%%%%%%%%%%%%%%%%%%%%%%%%%%%%%%%%%%%%%%%%%%%%%%%%%%%%%%%%%%%%%%%
\section{Introduction} \label{sec:introduction}

The Kerr spacetime remains one of the most significant solutions of general relativity, both from physical \cite{kw5g-d732} and mathematical \cite{Andersson.2022, Klainerman.2022, Dafermos.2024} perspectives. Beyond its well-established role as the exterior geometry of rotating black holes \cite{Kerr.1963}, its maximal analytic extension reveals a remarkably rich global structure comprising two asymptotically flat regions connected through the ring singularity \cite{carter_global_1968,Hawking.1973, chrusciel_geometry_2020}.
Null geodesics that traverse this singular ring offer a rare glimpse into the less-explored interior of the Kerr geometry, providing valuable insight into the causal structure, optical properties, and potential physical interpretation of the extended manifold.

In this work, we focus on a special class of trajectories--\textit{vortical null geodesics}--which cross from one asymptotically flat region ($r > 0$) to the other ($r < 0$). These geodesics are characterized by a negative Carter constant and exhibit no radial turning points. They thus connect regions that are causally separated by both event horizons and the disk enclosed by the singular ring. Our aim is twofold: first, to describe their properties in detail using both analytic and numerical methods; and second, to visualize how an observer in the negative-$r$ region would perceive light arriving from sources in the positive-$r$ region.

By recasting the constants of motion in terms of impact parameters, we parametrize the observer's celestial sphere and identify the subset corresponding to the \textit{inner throat} \cite{Helliwell.1975}, a domain in which the radial potential remains positive and geodesics traverse the extended spacetime without reflection. We show that only photons with impact parameters inside this inner throat reach an observer at $r_o = -\infty$ from sources at $r_s = +\infty$. Analysing the angular potential, we find at most two distinct constant-$\theta$ geodesics within this region, one corresponding to the principal null direction, and the other existing only for specific combinations of the rotation parameter and observer inclination. Furthermore, our analysis reveals the presence of a forbidden range of polar angles, a region of the source's sky that cannot be observed from the negative-$r$ domain.

Using these insights, we integrate the full geodesic equations in Eddington-Finkelstein-like horizon penetrating coordinates. The excellent agreement between the analytic and numerical solutions allows us to correct and refine previous expressions present in the literature. We then simulate the view of an observer located at $r_o = -\infty$, demonstrating that multiple images of the light source appear, clustering near the boundary of the inner throat. The sky seen by such an observer is found to be both distorted and flipped in azimuthal and polar directions. Finally, we note that our results also apply to a white hole configuration in which light sources are located at $r_s < 0$, suggesting that such geometries could, in principle, produce distinctive radiative signatures.

%%%%%%%%%%%%%%%%%%%%%%%%%%%%%%%%%%%%%%%%%%%%%%%%%%%%%%%%%%%%%%%%%%%%%%%%%%%%%%%%%%%%%
\section{The Kerr Metric} \label{sec:metric}

This section introduces the two coordinate systems for studying null geodesics in the Kerr spacetime \cite{Kerr.1963}. Additionally, key features of the Kerr solution that are essential for the subsequent analysis are discussed.

We use Eddington-Finkelstein-like coordinates $(u,r,\theta,\psi)$, based on the Boyer-Lindquist coordinate system, tailored explicitly to the principal null congruences of the Kerr solution, i.e., adapted to in- and outgoing null geodesics \cite{poisson_relativists_2004}.
They are particularly useful for solving the geodesic equations (see Sec.~\ref{sec:eom}).
Using these coordinates, the Kerr metric tensor is given by
\begin{gather}\label{eq:EF-metric}
    \begin{aligned}
        g=&-\left(1-\frac{2mr}{\Sigma}\right)\diff u^2-2\nu_r\,\diff r\,\diff u +\Sigma \,\diff \theta^2  \\
        & + \frac{(r^2+a^2)^2-\Delta a^2\sin^2\theta}{\Sigma} \sin^2\theta \,\diff \psi^2 \\
        & +2\nu_r a \sin^2\theta\,\diff \psi\,\diff r -\frac{4amr \sin^2\theta}{\Sigma}\,\diff \psi \,\diff u \,,
    \end{aligned}
\end{gather}
where $\nu_r = \pm 1$ and
\begin{align} \label{eq:sigma_delta}
    \Sigma = r^2+a^2\cos^2\theta \,, \quad \Delta = r^2 + a^2 - 2mr \,.
\end{align}
Depending on $\nu_r$, the coordinate system thus defined is based on the ingoing ($\nu_r = -1$) or outgoing ($\nu_r = +1$) congruence of the Kerr spacetime.
The constants $m$ and $a$ correspond to the mass and rotation of the black hole, respectively.

Furthermore, we need the so-called Kerr-Schild coordinates $(t,x,y,z)$, in which the metric tensor is given by
\begin{gather}\label{eq:KS-metric}
    \begin{aligned}
        g=\eta +\frac{2mr^3}{r^4+a^2z^2} \Bigg(&\diff t+\frac{r(x\,\diff x+y\,\diff y)}{a^2+r^2}\\
        &+\frac{a(y\,\diff x-x\,\diff y)}{a^2+r^2}+\frac{z}{r}\,\diff z \Bigg)^2 \,,
    \end{aligned}
\end{gather}
where $\eta = -\diff{t}^2+\diff{x}^2+\diff{y}^2+\diff{z}^2$ is the Minkowski metric.
We will utilize these coordinates for visualization purposes because of their flat background.
The transformation from \eqref{eq:EF-metric} to \eqref{eq:KS-metric} is given by
\begin{gather}\label{eq:trafo_EF_to_KS}
    \begin{aligned}
        x+iy &=(r+ia)\,e^{i(\psi+(1+\nu_r) r^\#)} \sin(\theta) \,,\\
        z &=r\cos(\theta) \,,\\
        \tilde{t} &=u+(1+\nu_r) r^*-r \,,
    \end{aligned}
\end{gather}
where we defined
\begin{align}\label{eq:r_hash_and_r_star} 
    r^*:=\int \frac{r^2+a^2}{\Delta} \diff r \,, \quad r^\#:=\int \frac{a}{\Delta} \diff r \,.
\end{align}
The asymptotic flatness of the Kerr spacetime at $r \rightarrow \pm \infty$ is evident from \eqref{eq:KS-metric}.
These two separate regions, which differ in the sign of the radial coordinate, are connected by the disk enclosed by the ring singularity, see below.

Three particular features of the Kerr spacetime are relevant to this work.
The first is the event horizons, given by $\{\Delta = 0 \Leftrightarrow r = m\pm \sqrt{m^2-a^2}\}$.
Horizons are present only for subcritical rotation ($0\leq a \leq m$); otherwise, the solution corresponds to a naked singularity.
Note that both the Eddington-Finkelstein and the Kerr-Schild coordinates are regular across the horizons.
A second feature is the ring singularity, defined by $\{\Sigma = 0 \Leftrightarrow r=0, \theta = \pi/2\}$, which is a curvature singularity \cite{chrusciel_structure_2020}.
The third is a causality-violating region \cite{chrusciel_geometry_2020}, also known as Carter's time machine, characterized by regions where the Killing vector $\partial_\psi$ is timelike, i.e., $g_{\psi \psi}<0$.
This region is generally not empty, and its name stems from the fact that a future-directed causal curve can connect any two points in this region.

%%%%%%%%%%%%%%%%%%%%%%%%%%%%%%%%%%%%%%%%%%%%%%%%%%%%%%%%%%%%%%%%%%%%%%%%%%%%%%%%%%%%%
\section{Geodesic Equations of Motion in the Kerr Spacetime} \label{sec:eom}

The geodesic equations in Eddington-Finkelstein coordinates for the wave vector $k^{\alpha}=(\dot{r},\dot{\theta},\dot{\phi},\dot{u})$, satisfying $k_{\alpha}k^{\alpha}=0$, take the form
\begin{align}
\Sigma\, \dot{r} &= \nu_r \sqrt{R(r)} \,,\label{eq:diff_eq_of_mot_EF_r}\\[4pt]
\Sigma\, \dot{\theta} &= \nu_\theta \sqrt{\Theta(\theta)} \,,\label{eq:diff_eq_of_mot_EF_theta}\\[4pt]
\Sigma\, \dot{\psi} &= \frac{a}{\Delta} \left(- \sqrt{R(r)}+ E (r^2+a^2)- a L \right) \nonumber\\
&\hspace{4mm}+ \frac{L}{\sin^2\theta} - aE  \,,\label{eq:diff_eq_of_mot_EF_psi} \\[4pt]
\Sigma\, \dot{u} &= \frac{r^2+a^2}{\Delta} \left(- \sqrt{R(r)}+ E (r^2+a^2)- a L  \right) \nonumber\\
&\hspace{4mm} +aL - a^2 E \sin^2 \theta \,,\label{eq:diff_eq_of_mot_EF_u}
\end{align}
where $\nu_r=\pm1$ and $\nu_\theta=\pm1$ denote the signs of $\dot{r}$ and $\dot{\theta}$ respectively, and the dot denotes differentiation with respect to an affine parameter $s$ along the geodesic. The functions $R(r)$ and $\Theta(\theta)$ are the radial and angular potentials,
\begin{align}
R(r) &= \left( E (r^2+a^2)- a L \right)^2 \nonumber\\
&\hspace{4mm} - \Delta \left(Q + (L-a E)^2 \right) \,, \label{eq:radial_pot}\\[2pt]
\Theta(\theta) &=Q-\cos^2\theta \left(-E^2 a^2+\frac{L^2}{\sin^2\theta}\right) \,.\label{eq:polar_pot}
\end{align}
The above equations contain the constants of motion corresponding to the energy $E = - k_u$, the angular momentum around the axis of symmetry $L = k_\psi$, and the redefined Carter's constant \cite{carter_global_1968}
\begin{equation} \label{eq:carter_constant1}
Q = k_\theta^2 + \cos^2\theta \left( -E^2 a^2 + \frac{L^2}{\sin^2\theta} \right) \,.
\end{equation}
For null geodesics, we can work with
\begin{align} \label{eq:const_lambda_eta}
    \lambda = \frac{L}{E} \,, \quad \eta = \frac{Q}{E^2}
\end{align}
instead, meaning that only the sign of $E$ is important.
We restrict the value of $E$ as follows:
First, since we want to investigate geodesics that reach the region where $\partial_u$ is timelike, we can assume $E>0$ \cite{oneill_geometry_2014}.
Second, we can parameterize the geodesic so that the energy is normalized $E=1$, and thus $\lambda$ is the azimuthal angular momentum of the geodesic.
Consequently, we will set $E=1$ in any numerical calculations.

In addition to the new constants of motion in \eqref{eq:const_lambda_eta}, we take advantage of a cleverly chosen parameter $\tau$ along the geodesic, which makes it easier to solve the equations of motion for null geodesics.
The so-called Mino time \cite{mino_perturbative_2003} is defined by
\begin{equation} \label{eq:mino_time}
\diff \tau =\frac{E}{\Sigma}\, \diff s \,.
\end{equation}
Rewriting \eqref{eq:diff_eq_of_mot_EF_r}-\eqref{eq:diff_eq_of_mot_EF_u} using \eqref{eq:const_lambda_eta} and \eqref{eq:mino_time}, as well as $\mu=0$, we arrive at the set of differential equations we aim to solve:
\begin{align}
    \frac{\diff r}{\diff \tau} &= \nu_r \sqrt{R(r)} \,, \label{eq:diff_eq_of_mot_EF_Mino_r}\\[3pt]
    \frac{\diff \theta}{\diff \tau} &= \nu_\theta \sqrt{\Theta(\theta)} \,, \label{eq:diff_eq_of_mot_EF_Mino_theta}\\[3pt]
    \frac{\diff \psi}{\diff \tau} &= \frac{a}{\Delta} (2 m r -a \lambda - \sqrt{R(r)}) +\frac{\lambda}{\sin^2\theta} \,, \label{eq:diff_eq_of_mot_EF_Mino_psi}\\[3pt] 
    \frac{\diff u}{\diff \tau} &= \frac{r^2+a^2}{\Delta} (r^2+a^2-a \lambda - \sqrt{R(r)}) \nonumber\\
    &\hspace{4mm} + a(\lambda - a \sin^2\theta) \,, \label{eq:diff_eq_of_mot_EF_Mino_u}
\end{align}
with new radial and angular potentials
\begin{align}
    R(r) &= (r^2+a^2-a \lambda)^2-\Delta (\eta+(\lambda -a)^2) \,, \label{eq:rad_pot_null}\\[2pt]
    \Theta(\theta) &= \eta + a^2 \cos^2\theta-\lambda^2 \frac{\cos^2\theta}{\sin^2\theta} \,.  \label{eq:ang_pot_null}
\end{align}

%%%%%%%%%%%%%%%%%%%%%%%%%%%%%%%%%%%%%%%%%%%%%%%%%%%%%%%%%%%%%%%%%%%%%%%%%%%%%%%%%%%%%
\section{Visualization preliminaries}

We want to study geodesics that start at some source at a positive radius $r_s > 0$ and reach the observer located at $r_o=-\infty$, hence we consider ingoing null geodesics.
We want to emphasize that the light sources we consider throughout this paper are test sources that do not alter the spacetime or its symmetries.
As such, they can be mathematically identified with points within the spacetime.

%%%%%%%%%%%%%%%%%%%%%%%%%%%%%%%%%%%%%%%%%%%%%%%%%%%%%%%%%%%%%%%%%%%%%%%%%%%%%%%%%%%%%%%%%%%%%%%%
\subsection{Radial Motion}
A null geodesics' radial motion is determined by \eqref{eq:diff_eq_of_mot_EF_Mino_r}, which, upon inserting \eqref{eq:sigma_delta} and \eqref{eq:rad_pot_null}, becomes
\begin{multline} \label{eq:rad_eq_EF_full}
\frac{\diff r}{\diff \tau} = \nu_r \Big\{ (r^2+a^2-a \lambda)^2
\\
-(r^2+a^2-2mr) (\eta+(\lambda -a)^2)\Big\}^{1/2} \,.
\end{multline}
It should be clear that the radial motion for physical geodesics is restricted to the case where the radicand in \eqref{eq:rad_eq_EF_full} is nonnegative.
Therefore, the real-valued roots of $R(r)$ determine the number of radial turning points of a null geodesic.
For our study, we are interested in geodesics connecting a spacetime point with $r_s\gg0$ to another spacetime point with $r_o\ll0$.
Such geodesics must consequently have zero radial turning points (cf. \cite{oneill_geometry_2014}), i.e., $R(r)\neq 0$ for all $r$.
Only if $\eta < 0$, \eqref{eq:rad_eq_EF_full} can exhibit four complex roots and, therefore, no radial turning point \cite{gralla_null_2020}.
Geodesics with negative $\eta$ are also known as \textit{vortical null geodesics}, and they are the main focus of our study.

%%%%%%%%%%%%%%%%%%%%%%%%%%%%%%%%%%%%%%%%%%%%%%%%%%%%%%%%%%%%%%%%%%%%%%%%%%%%%%%%%%%%%%%%%%%%%%%%
\subsubsection{Impact parameters}
Consider an observer located at $(u_o, r_o, \theta_o, \psi_o)$ and a light source at $(u_s, r_s, \theta_s, \psi_s)$.\footnote{Throughout this paper, the subscripts $o$ and $s$ indicate the coordinates corresponding to the spacetime location of an observer or a source, respectively.}
For $|r_o|\rightarrow\infty$, it is helpful to replace $\lambda$ and $\eta$ in \eqref{eq:const_lambda_eta} with the so-called impact parameters $\alpha$ and $\beta$, see \cite{bardeen_timelike_1972}.
Based on the assumption that an observer far away from the black hole measures the direction of photons that reach them relative to the center of symmetry of the spacetime (i.e., the observer's line of sight is towards the center of the black hole), we define\footnote{A detailed derivation of these impact parameters can be found in \cite{bardeen_timelike_1972}.
The only difference to \eqref{eq:impact_param_alpha_beta} is an overall sign in the equation for $\alpha$.
This stems from the fact that we consider $r_o <0$.}
\begin{gather} \label{eq:impact_param_alpha_beta}
    \begin{aligned}
        \alpha &= \frac{\lambda}{\sin\theta_o} \,, \\
        \beta &= \nu_\theta \sqrt{\eta +a^2 \cos^2\theta_o - \lambda^2\, \frac{\cos^2\theta_o}{\sin^2\theta_o}} \, .
    \end{aligned}
\end{gather}

For an observer substantially far away from the black hole, these impact parameters serve as a coordinate grid on a small patch of the sky centered around the black hole.
Therefore, $\alpha$ and $\beta$ are coordinates in the field of view for an observer looking directly at the black hole.
In this case, $\alpha$ and $\beta$ describe the displacements perpendicular and parallel to the axis of symmetry, respectively.

Note that $\alpha$ and $\beta$ depend on the observer's location; hence, the analysis of the radial potential becomes observer-dependent.
Eq.~\eqref{eq:impact_param_alpha_beta} causes problems only at $\sin\theta_o=0$, corresponding to an observer located on the axis of symmetry.
However, any geodesic going through the axis of symmetry (at which $\sin\theta=0$), and therefore especially those reaching an observer located in the axis of symmetry (with $\theta_o=0$), must necessarily have $\lambda=0$ (see, e.g., \cite{oneill_geometry_2014}).
The statement $\sin\theta_o=0 \Leftrightarrow \lambda=0$ also becomes apparent when considering the inverses of \eqref{eq:impact_param_alpha_beta}:
\begin{align} \label{eq:impact_param_alpha_beta_inverse}
\lambda = \alpha \sin\theta_o \, , \quad \eta = \beta^2+(\alpha^2-a^2)\cos^2\theta_o \, .
\end{align}

%%%%%%%%%%%%%%%%%%%%%%%%%%%%%%%%%%%%%%%%%%%%%%%%%%%%%%%%%%%%%%%%%%%%%%%%%%%%%%%%%%%%%%%%%%%%%%%%
\subsubsection{Inner throat}
Expressing \eqref{eq:rad_pot_null} in terms of $\alpha$ and $\beta$ yields
\begin{multline}
        R(r) = (r^2+a^2-a\,\alpha\,\sin\theta_o)^2 - (r^2-2mr+a^2) \\
         \times\left( \beta^2 + (\alpha^2 - a^2) \,\cos^2\theta_o + ( \alpha \,\sin\theta_o - a)^2 \right) \,.
\end{multline}
As a quartic function in $r$, it has four complex roots.
The conditions $R(r)=0$ and $\diff R(r)/\diff r=0$ define two non-intersecting closed curves in the $(\alpha,\beta)$-plane, demarcating regions with a different number of roots of $R(r)$:
Within the inner region, $R(r)$ has zero real roots; between the inner and outer curves, it has two real roots and two complex roots; outside the outer curve, it has four real roots.
The inner region is known as the inner throat, and we use the term ``a geodesic within the inner throat'' as a shorthand for ``a geodesic with impact parameters within the inner throat''.

\begin{figure}[t]\centering
\includegraphics[width=1.0\columnwidth, keepaspectratio]{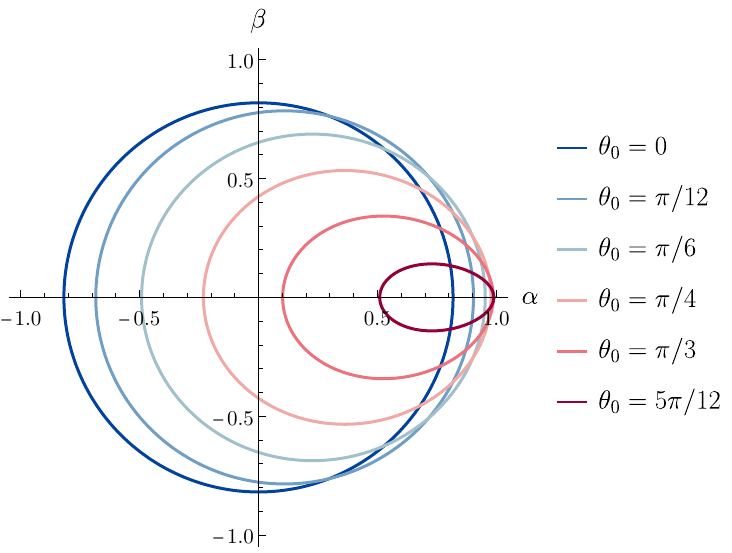}
\caption[Inner throats for various $\theta_o$]{
The inner throat for different polar angles of an observer $\theta_o$ and rotational parameter $a/m=0.99$.
The case $\theta_o=0 \, (\pi/2)$ corresponds to an observer located on the axis of symmetry (equatorial plane).
For $\theta_o=\pi/2$, the inner throat vanishes.
}
\label{fig:inner_throats}
\end{figure}

As mentioned above, there are no turning points in the radial coordinate for geodesics within the inner throat, as these coincide with the real roots of $R(r)$.
Thus, only null geodesics within the inner throat connect a light source at $r_s \gg 0$ and an observer at $r_o\ll 0$. 
Therefore, the inner throat serves as the field of view for our observer at $r_o\ll 0$ with a light source at $r_s\gg 0$.
How the inner throat changes depending on the observer's polar coordinate is depicted in Fig.~\ref{fig:inner_throats}.
In particular, the left/right vertices, i.e., the inner throat boundary for $\beta = 0$ and minimal/maximal $\alpha$, move to the right and become closer to each other when $\theta_o$ increases from $0$ to $\pi/2$, cf. \cite{Helliwell.1975}.

%%%%%%%%%%%%%%%%%%%%%%%%%%%%%%%%%%%%%%%%%%%%%%%%%%%%%%%%%%%%%%%%%%%%%%%%%%%%%%%%%%%%%%%%%%%%%%%%
\subsection{Polar motion} \label{subsec:polar_motion}
The polar motion of a null geodesic in Kerr spacetime is governed by \eqref{eq:diff_eq_of_mot_EF_Mino_theta}, which, upon inserting \eqref{eq:ang_pot_null}, becomes
\begin{equation} \label{eq:diff_eq_of_mot_EF_Mino_theta_full}
\frac{\diff \theta}{\diff \tau} = \nu_\theta \sqrt{\eta + a^2 \cos^2\theta-\lambda^2 \frac{\cos^2\theta}{\sin^2\theta}} \,.
\end{equation}
A feature of vortical null geodesics ($\eta <0$) is that they are bound to one hemisphere, i.e., the geodesic is trapped either in the northern hemisphere $0\leq\theta<\pi/2$ or in the southern hemisphere $\pi/2<\theta\leq\pi$.
The radicand in \eqref{eq:diff_eq_of_mot_EF_Mino_theta_full} is generally non-zero, so the geodesic exhibits oscillatory behavior in the polar coordinate.
The polar turning points are given by \cite{gralla_null_2020}
\begin{equation} \label{eq:theta_pm}
\theta_\pm=\arccos(h \sqrt{w_\mp}) \,,
\end{equation}
where $h=\mathrm{sign}(\cos\theta)$ determines the hemisphere, and
\begin{equation} \label{eq:u_pm}
w_\mp = \frac{a^2-\eta-\lambda^2}{2a^2} \mp \sqrt{\left( \frac{a^2-\eta-\lambda^2}{2a^2} \right)^2 + \frac{\eta}{a^2}} \,.
\end{equation}
Inserting \eqref{eq:impact_param_alpha_beta_inverse} into \eqref{eq:theta_pm}-\eqref{eq:u_pm} yields $\theta_\pm(\alpha,\beta,\theta_o)$ with which we can further analyze the polar motion for geodesics inside the inner throat.

%%%%%%%%%%%%%%%%%%%%%%%%%%%%%%%%%%%%%%%%%%%%%%%%%%%%%%%%%%%%%%%%%%%%%%%%%%%%%%%%%%%%%%%%%%%%%%%%
\subsubsection{Geodesics with constant polar angle and principal null congruence} \label{sec:theta_principal_null_congruence}
By solving the equation $w_+=w_-$ we get points in the impact-parameter-plane where geodesics have constant polar angle $\theta=\theta_+=\theta_-$.
This yields four solutions for $\beta(\alpha)$
\begin{equation} \label{eq:theta_pm_beta}
\beta = \pm\sqrt{- (\alpha \pm a\sin\theta_o)^2} \,.
\end{equation}
The parameter $\beta$ is real only when the radicand on the right-hand side is zero.
Thus, the only two points in the $(\alpha,\beta)$-space with constant $\theta$ along the geodesic are
\begin{equation}\label{eq:alpha_beta_for_const_theta}
(\alpha_\pm , \beta_0) = (\pm a\sin\theta_o , 0) \,.
\end{equation}

To determine if an observer with $r_o\ll 0$ receives photons with a constant polar angle originating at $r_s\gg 0$, it remains to determine if $(\alpha_\pm , \beta_0)$ lies inside the inner throat.
For this, we take the parametric solution for the inner throat boundary obtained by solving $R(r)=\diff R(r)/\diff r=0$ for $(\alpha, \beta)$ and compute for which $a$ and $\theta_o$ this parametric curve coincides with \eqref{eq:alpha_beta_for_const_theta}.
This results in eight solutions for the pair $(\alpha_+,\beta_0)$,
\begin{equation}\label{eq:param_sol_theta_const_at_bnd_right}
\begin{aligned}
a &= \pm \sqrt{r(2m-r)}\,,\\
\theta_o &= \pm\arccos\left( \pm\sqrt{\frac{r}{r-2m}} \right)\,,
\end{aligned}
\end{equation}
and eight solutions for $(\alpha_-, \beta_0)$,
\begin{equation}\label{eq:param_sol_theta_const_at_bnd_left}
\begin{aligned}
a &= \pm \sqrt{\frac{r^2(3m-2r)}{m}}\,, \\
\theta_o &= \pm \arccos\left( \pm \sqrt{\frac{3m}{3m-2r}} \right)\,.
\end{aligned}
\end{equation}

Eq.~\eqref{eq:param_sol_theta_const_at_bnd_right} has no real-valued solutions for $a$ and $\theta_o$.
To see this, we calculate $a(\theta_o)$ by expressing $r$ in terms of $\theta_o$ using the second equation of \eqref{eq:param_sol_theta_const_at_bnd_right} and substituting this into the first equation, resulting in
\begin{align}\label{eq:a_theta_pm_left_plus}
a_{crit, +}(\theta_o) = \pm 2 i m \frac{|\cot\theta_o|}{\sin\theta_o} \,.
\end{align}
This does not have real-valued solutions except for $a_{crit,+}= 0$ and $\theta_o= \pi/2$.
Thus, $(\alpha_+,\beta_0)$ never crosses the inner throat boundary.
Because for $a/m=0.99$ and $\theta_o=0$, the point $(\alpha_+,\beta_0)$ is inside the inner throat (see Fig.~\ref{fig:inner_throats}), we conclude that these geodesics are always inside the inner throat.
Therefore, every observer at $r_o\ll 0$ with a polar angle $\theta_o\neq\pi/2$ receives photons with a constant polar angle at the impact parameters $(\alpha_+ ,\, \beta_0)$ originating at $r_s>0$.
These geodesics correspond to the principal null congruence on which the Eddington-Finkelstein coordinates \eqref{eq:EF-metric} are based.
As such, they also have constant $\psi$ and $u$ (see \cite{poisson_relativists_2004} or Appendix~\ref{appsec:principal_null_congruence_psi_and_u}).

\begin{figure}[t!]\centering
\includegraphics[width=1.0\columnwidth, keepaspectratio]{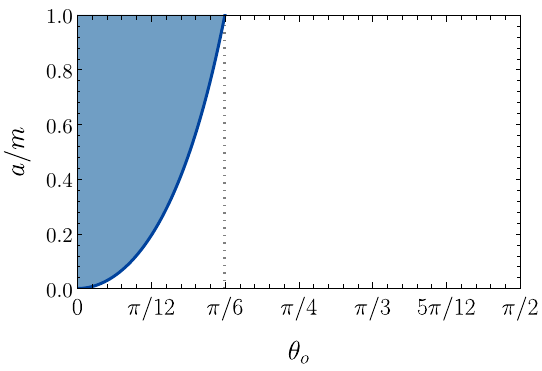}
\caption[$a$-$\theta_o$-plot for constant polar angle at right edge of inner throat]{\label{fig:a_for_theta_pm_left} Plot of \eqref{eq:a_theta_pm_left_minus} for $0\leq \theta_o<\pi/2$. The shaded area indicates combinations of $a/m$ and $\theta_o$ where two distinct geodesics with constant $\theta$ lie inside the inner throat. On the boundary curve, the left vertex of the inner throat exhibits a constant polar angle, corresponding to a geodesic with a constant polar angle captured in an orbit around the black hole. The vertical dotted line corresponds to $\theta_o=\pi/6$.}
\end{figure}

Similarly, we solve \eqref{eq:param_sol_theta_const_at_bnd_left} for $a(\theta_o)$ and obtain
\begin{align} \label{eq:a_theta_pm_left_minus}
a_{crit,-}(\theta_o) = \frac{3\sqrt{3}m}{2} \frac{\tan^2\theta_o}{|\cos\theta_o|} \,,
\end{align}
where we chose the positive sign for the square root as we restrict ourselves to the case $a/m>0$.
For $a>a_{crit,-}$, an observer at $\theta_o$ perceives two distinct points with a constant polar angle inside the inner throat.
Fig.~\ref{fig:a_for_theta_pm_left} shows the region $a>a_{crit_-}$ for $0\leq\theta_o\leq\pi/2$.
Since the physically acceptable cases are restricted to $a<m$, only a limited range of observer positions allows for the existence of two constant-$\theta$ geodesics traversing the inner throat.
Solving $a_{crit,-}/m=1$ using \eqref{eq:a_theta_pm_left_minus} gives $\theta_{o,max}=\pi/6$ as an upper bound.
Consequently, only observers with $0<\theta_o<\pi/6$ potentially observe these two distinct geodesics, depending on the rotation speed $a$ of the black hole.
However, as the resulting geodesics do not correspond to principal null directions, $\psi$ and $u$ are not constant along geodesics with $(\alpha_-,\beta_0)$.

%%%%%%%%%%%%%%%%%%%%%%%%%%%%%%%%%%%%%%%%%%%%%%%%%%%%%%%%%%%%%%%%%%%%%%%%%%%%%%%%%%%%%%%%%%%%%%%%
\subsubsection{Geodesic motion close to the axis of symmetry or the equatorial plane} \label{sec:polar_angle_close_to_equator_or_axis_of_symmetry}
From the behavior of $\mathrm{arccosine}$ and \eqref{eq:theta_pm}, we conclude that the polar turning point closest to the axis of symmetry (equatorial plane) is $\theta_-$ ($\theta_+$), regardless of the sign of $h$ (hemisphere).
By substituting $\alpha=0$, we find
\begin{align}
\restr{\theta_-}{\alpha=0} = \arccos (h) = \begin{cases}
0\, , \quad h=+1\\
\pi\, , \quad h=-1\\
\end{cases} \, .
\end{align}
Consequently, geodesics with $\alpha=0$ reach the axis of symmetry.
This fact will be crucial in Sec.~\ref{subsec:theta-sol} to correct the analytical solution for $\theta$ and obtain a smooth solution for every null geodesic inside the inner throat.

\begin{figure}[t!]\centering
\includegraphics[width=1.0\columnwidth, keepaspectratio]{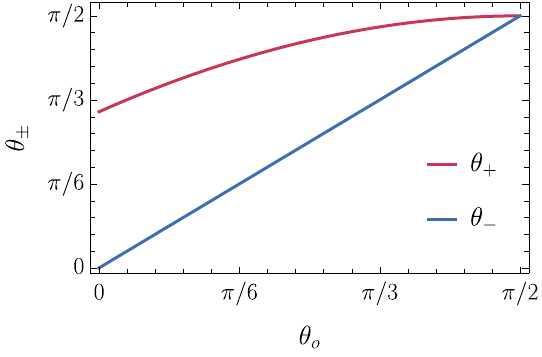}
\caption[$\theta_\pm$ at the left vertex of the inner throat]{\label{fig:theta_pm_right_edge} The angles $\theta_\pm$ for the right vertex ($\alpha=\alpha_{\textrm{max}}$, $\beta=0$) of the inner throat for varying polar angle of the observer $\theta_o$. The rotation parameter is set to $a/m=0.99$.}
\end{figure}

On the other hand, from \eqref{eq:theta_pm} and \eqref{eq:u_pm} together with \eqref{eq:impact_param_alpha_beta_inverse}, we find that inside the inner throat, the point closest to the equatorial plane (with $\theta_{+,\max}$) is reached at the right boundary of the throat, so for $(\alpha_{\max},\beta=0)$ \cite{noauthor_httpswwwquantagonatmasters-thesis_nodate}.
As $\theta_{+,\max}<\pi/2$ for $h=1$ and $\theta_{+,\max}>\pi/2$ for $h=-1$, there is a ``forbidden'' region for the polar motion of vortical null geodesics inside the inner throat: they only reach values satisfying $0<\theta_- \leq \theta \leq \theta_+ \leq \theta_{+,\max} < \pi/2$ for $h=1$ or $\pi/2 <\theta_{+,\max} \leq \theta_+ \leq \theta \leq \theta_-<\pi$ for $h=-1$.
Fig.~\ref{fig:theta_pm_right_edge} shows $\theta_\pm$ for the right vertex of the inner throat with $(\alpha_{\max},\beta=0)$ and $h=1$.
For a given $\theta_o$, polar angles greater than $\theta_+$ constitute the forbidden polar angle region.

%%%%%%%%%%%%%%%%%%%%%%%%%%%%%%%%%%%%%%%%%%%%%%%%%%%%%%%%%%%%%%%%%%%%%%%%%%%%%%%%%%%%%
\section{Solving the geodesic equations} \label{chapter:solving_geod_eq}
In this section, we solve the geodesic equations for vortical null geodesics. Analytic solutions are derived in Sec.~\ref{sec:ana-sol}, while in Sec.~\ref{sec:num-sol} the corresponding integrals are evaluated numerically to verify the analytic results.

To solve the equations of motion \eqref{eq:diff_eq_of_mot_EF_Mino_r}-\eqref{eq:diff_eq_of_mot_EF_Mino_u}, a suitable approach is to cast them in integral form:
\begin{widetext}
\begin{gather}
\tau_{o}-\tau_{s} = \int_{\theta_s}^{\theta_o}\frac{\diff \theta}{\nu_\theta \sqrt{\Theta(\theta)}} =\, \int_{r_s}^{r_o}\frac{\diff r}{\nu_r \sqrt{R(r)}} \,, \label{eq:int_eq_of_mot_EF_Mino_r_theta}\\
\psi_o - \psi_s =\, \nu_r \int_{r_s}^{r_o} \frac{a(2mr-a\lambda -\sqrt{R(r)})}{\Delta \sqrt{R(r)}} \,\diff r \,+ \nu_\theta \int_{\theta_s}^{\theta_o} \frac{\lambda}{\sin^2(\theta) \sqrt{\Theta(\theta)}} \,\diff \theta \,,\label{eq:int_eq_of_mot_EF_Mino_psi} \\ 
u_o - u_s =\, \nu_r \int_{r_s}^{r_o} \frac{r^2 \Delta +2mr(r^2+a^2-a\lambda)-(r^2+a^2)\sqrt{R(r)}}{\Delta \sqrt{R(r)}} \,\diff r \,+ \nu_\theta \, a^2 \int_{\theta_s}^{\theta_o} \frac{\cos^2\theta}{\sqrt{\Theta(\theta)}} \,\diff \theta \,. \label{eq:int_eq_of_mot_EF_Mino_u}
\end{gather}
\end{widetext}
Eq.~\eqref{eq:int_eq_of_mot_EF_Mino_r_theta} couples the $r$- and $\theta$-motion, and both integrals are equal to the elapsed Mino time $\tau$ along the geodesic.
This approach is also taken in \cite{gralla_null_2020}, whose results we use whenever applicable.
However, at some points, we use a different approach, which we indicate accordingly.
The main difference in their calculations is that we work in Eddington-Finkelstein-like coordinates, which are globally regular.

Our analytical solutions are valid for the subclass of vortical null geodesics with no radial turning points.
Accordingly, the radial potential $R(r)$ never vanishes along the geodesic; hence, all its roots are complex.
Consequently, for the roots $r_i$ of $R(r)$, we can arrange that $r_1=r_2^*$ and $r_3=r_4^*$ (where the asterisk denotes complex conjugation).

Lastly, a few words about the general approach of calculating the starting values of a null geodesic that reaches the observer with impact parameters $(\alpha, \beta)$.
The fact that we fix the observer's position and the impact parameters leads us to calculate the geodesic's trajectory ``back in time.''
Thus, we calculate the trajectory of a vortical null geodesic originating from the observer at $r_o=-\infty$ and some $(\alpha, \beta)$ within the inner throat and terminating at the source point at $r_s>0$.
This technical inversion of the geodesic direction is equivalent to exchanging the integration boundaries.
However, we emphasize that these geodesics still correspond to ingoing rays, crossing from the exterior region of the Kerr black hole to the interior.
This is a standard approach in many calculations done for black hole visualizations (see, e.g., \cite{james_gravitational_2015, gralla_lensing_2020}).

%%%%%%%%%%%%%%%%%%%%%%%%%%%%%%%%%%%%%%%%%%%%%%%%%%%%%%%%%%%%%%%%%%%%%%%%%%%%%%%%%%%%%%%%%%%%%
\subsection{Analytic solutions} \label{sec:ana-sol}
To follow the general approach of \cite{gralla_null_2020} for the calculation of the antiderivatives of \eqref{eq:int_eq_of_mot_EF_Mino_r_theta}-\eqref{eq:int_eq_of_mot_EF_Mino_u}, we need to define some additional quantities, which can be found in Appendix~\ref{appsec:roots_and_definitions}.
Included are formulas for the roots $r_i$ ($i=1,2,3,4$) of the radial potential $R(r)$, as well as for the quantities $a_1$, $a_2$, $b_1$, $b_2$, $C$, $D$, $g_0$, $k$, $k_4$, $x_4(r)$, $h$, and $X_4(\tau)$.
We also define the integrals
\begin{align}
I_\pm &:= \nu_r \int_{r_s}^{r_o} \frac{ \diff r}{(r-r_\pm) \sqrt{R(r)}} \, ,\label{eq:I_pm}\\
I_\ell &:= \nu_r \int_{r_s}^{r_o} \frac{ r^\ell\, \diff r}{\sqrt{R(r)}} \,. \label{eq:I_l}
\end{align}
In the following, $\mathcal{I}$ will always denote the antiderivative of the corresponding integral $I$, so e.g. $\mathcal{I}_\pm$ is the antiderivative of $I_\pm$, i.e., $I_\pm = \mathcal{I}_\pm (r_o) - \mathcal{I}_\pm (r_s)$.
The primary method to solve the equations is to reduce them to elliptic integrals.
A short introduction, including the relevant elliptic functions, can be found in Appendix~\ref{subsec:elliptic_integrals}.

%%%%%%%%%%%%%%%%%%%%%%%%%%%%%%%%%%%%%%%%%%%%%%%%%%%%%%%%%%%%%%%%%%%%%%%%%%%%%%%%%%%%%%%%%%%%%
\subsubsection{\texorpdfstring{$\tau$}{tau}-integral} \label{subsec:tau-sol}
To get a solution for the Mino time along the geodesic, we rewrite \eqref{eq:diff_eq_of_mot_EF_Mino_r} as
\begin{align} \label{eq:tau_eq_1}
\diff \tau = \nu_r \frac{\diff r}{\sqrt{R(r)}} \,.
\end{align}
By integrating from the source to the observer, we obtain
\begin{align} \label{eq:tau_eq_2}
\tau_o - \tau_s = \nu_r \int_{r_s}^{r_o} \frac{\diff r}{\sqrt{R(r)}} \, .
\end{align}
The right-hand side equals $I_0$ in \eqref{eq:I_l} and according to \cite{gradshtein_table_2000}~(§3.145), its antiderivative is
\begin{align} \label{eq:antider_I0}
\mathcal{I}_0(r) = \frac{2 \nu_r}{C+D} \, F\left(\arctan(x_4(r))+\arctan(g_0),\, k_4\right) \, ,
\end{align}
and we find
\begin{align} \label{eq:sol_tau}
\tau_o = \tau_s + \mathcal{I}_0(r_o) - \mathcal{I}_0(r_s) \,,
\end{align}
which is the solution for the Mino time $\tau_o$ along the geodesic depending on the radial position of the source $r_s$ and of the observer $r_o$ as well as the initial Mino time $\tau_s$ at the source.
Setting $\tau_s=0$ yields the elapsed Mino time along the geodesic.

%%%%%%%%%%%%%%%%%%%%%%%%%%%%%%%%%%%%%%%%%%%%%%%%%%%%%%%%%%%%%%%%%%%%%%%%%%%%%%%%%%%%%%%%%%%%%
\subsubsection{\texorpdfstring{$r$}{r}-integral} \label{subsec:r-sol}
\begin{figure}[t]
\centering
\includegraphics[width=1.0\columnwidth, keepaspectratio]{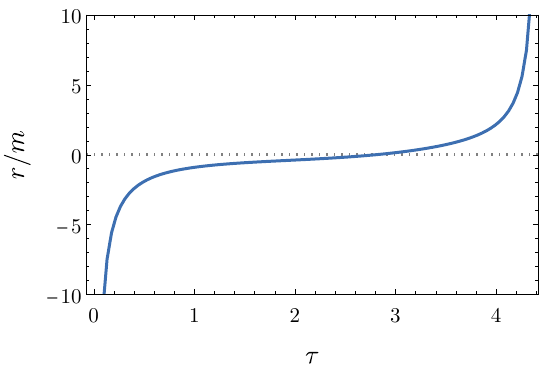}
  \caption[Example radial trajectory]{\label{fig:example_traj_r_combined} An example radial trajectory of a vortical null geodesic with impact parameters $\alpha=-0.15$ and $\beta=0.1$ within the inner throat for $\theta_o=\pi/4$. The black hole rotation parameter is set to $a/m=0.99$. Because we measure distances in terms of the black hole's mass $m$, we are plotting $r/m$.}
\end{figure}

To obtain a formula for the radial component of a vortical null geodesic, the authors of \cite{gralla_null_2020} inverted \eqref{eq:sol_tau} by using basic properties of elliptic integrals and elliptic functions to derive a formula for $r_o(\tau)$ (see their Sec.~4 in Appendix~B).
The final formula is
\begin{align} \label{eq:sol_r}
r_o(\tau) = -a_2 \left( \frac{g_0-\mathrm{sc}\left(X_4(\tau),k_4\right)}{1+g_0 \,\mathrm{sc}\left(X_4(\tau),k_4\right)} \right) -b_1 \,,
\end{align}
which depends implicitly on $r_s$ and $\nu_r$ via $X_4(\tau)$.
Using \eqref{eq:sol_tau}, we can first calculate the elapsed Mino time along the geodesic, while with \eqref{eq:sol_r}, we can track the geodesic along its radial path of increasing Mino time.
An example trajectory with $r_s=+\infty$, $r_o=-\infty$ and $\theta_o=\pi/4$ can be found in Fig.\,\ref{fig:example_traj_r_combined}.

%%%%%%%%%%%%%%%%%%%%%%%%%%%%%%%%%%%%%%%%%%%%%%%%%%%%%%%%%%%%%%%%%%%%%%%%%%%%%%%%%%%%%%%%%%%%%
\subsubsection{\texorpdfstring{$\theta$}{theta}-integral} \label{subsec:theta-sol}
With the solution for the polar angle, we proceed similarly to the previous subsection and state the final formula derived in \cite{gralla_null_2020} (the details can be found in Sec.~III~B there).
The polar angle along the geodesic is given by
\begin{widetext}
\begin{align} \label{eq:theta_sol_uncorrected}
\theta_o(\tau) = \arccos\left(h_s \sqrt{\gamma_-}\, \dn\left( \sqrt{\gamma_- \,a^2} \, \bigl(\tau + \nu_\theta \, \mathcal{G}_\theta(\theta_s)\bigr),\, 1-\frac{\gamma_+}{\gamma_-} \right) \right) \,,
\end{align}
where $h_s=\mathrm{sign}(\cos\theta_s)$, and
\begin{align}
    \gamma_\pm&=\frac{1}{4} \Biggl( 3-\frac{2(\alpha^2+\beta^2)}{a^2} + \cos(2\theta_o) \pm 4\Big\{\frac{(a^2(\cos(2\theta_o)+3)-2(\alpha^2+\beta^2))^2}{16a^4}
    +\frac{\beta^2 + \cos^2\theta_o (\alpha^2-a^2)}{a^2} \Big\}^{1/2} \Biggr) \,, \label{eq:u_pm_alpha_beta}\\
    \mathcal{G}_\theta(\theta) &= -\frac{h_s}{\sqrt{\gamma_- \,a^2}} \, F\left( \arcsin \sqrt{\frac{\cos^2\theta - \gamma_-}{\gamma_+ - \gamma_-}}  ,\, 1-\frac{\gamma_+}{\gamma_-} \right) \, . \label{eq:mathcal_G_theta}
\end{align}
\end{widetext}
Using this formula, we can calculate the $\theta$-trajectory of any geodesic with impact parameters inside the inner throat.
We can use the general periodicity properties of the elliptic function presented in Appendix~\ref{subsec:elliptic_integrals} to determine the period of the polar angle.

\begin{figure*}[ht]
  \centering
  \captionsetup[subfigure]{margin={1.5cm,0cm}}%
  \subfloat[][\label{subfig:ex_traj_theta_a} $\alpha=-0.15$, $\beta=0.1$]{\includegraphics[width=0.475\textwidth]{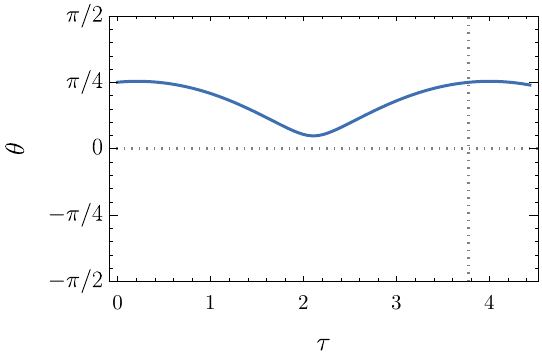}}
  \qquad
  \subfloat[][\label{subfig:ex_traj_theta_b} $\alpha=0$, $\beta=0.1$]{\includegraphics[width=0.475\textwidth]{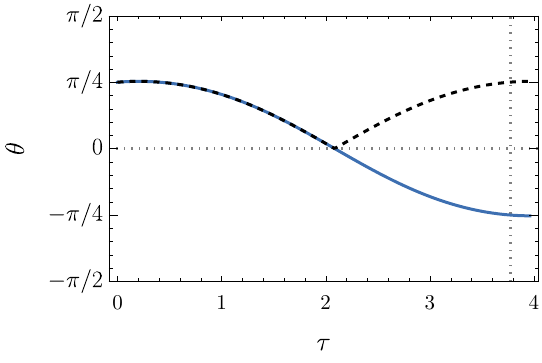}}
  \caption[Example polar angle trajectory (not smooth everywhere)]{
  \label{fig:example_traj_theta_combined}
  Two example $\theta$-trajectories calculated with \eqref{eq:theta_sol_uncorrected} with $r_s=+\infty$, $r_o=-\infty$, $\theta_o=\pi/4$, and $a/m=0.99$.
  \textbf{(a):} The solution is smooth in the general case $\alpha\neq0$. The vertical line indicates one full period of $\theta$.
  \textbf{(b):} For $\alpha=0$, the trajectory is not smooth at $\theta_-=0$ (black, dashed). As discussed in the main text, we obtain a smooth solution by extending it to negative $\theta$ (blue, solid). The hemisphere defined by $\mathrm{sign}(\cos\theta)$ is unaffected by this. The vertical line denotes one half-period of the polar angle.}%
\end{figure*}

However, there is a problem with the smoothness of geodesics with $\alpha=0$ (see Fig.~\ref{subfig:ex_traj_theta_b}).
For $\alpha\neq0$, the geodesic oscillates between $\theta_+$ and $\theta_-$.
The discussion in Sec.~\ref{sec:polar_angle_close_to_equator_or_axis_of_symmetry} shows that $\theta_-=0$ for $\alpha=0$, i.e., such geodesics reach the axis of symmetry.
Consequently, by approaching $\alpha=0$ in the impact parameter space, the turning point at $\theta_-$ loses its smooth nature and develops into a cusp.
To ensure a smooth solution throughout, we extend the solution to negative values of $\theta$ for $\alpha=0$, so that it oscillates in the range $-\theta_+ \leq \theta \leq \theta_+$.
This is achieved by flipping the sign of the right-hand side in \eqref{eq:theta_sol_uncorrected} for specific ranges of $\tau$, which are determined by the periodicity properties of the elliptic function $\dn(u,m)$.
The altered version of \eqref{eq:theta_sol_uncorrected} is still the antiderivative to \eqref{eq:diff_eq_of_mot_EF_Mino_theta} because of the ambiguity in the inverse of cosine (which is used in the derivation).

To implement this modification, we calculate the Mino time $\tau$ for which the right-hand side of \eqref{eq:theta_sol_uncorrected} becomes zero.
Solving $\theta_o(\tau)=0$ for $\tau$ yields
\begin{gather}\label{eq:tau_is_zero_1}
    \begin{aligned}
        \tau_0 =& \frac{1}{a\sqrt{\gamma_-}} \,F\left( \arcsin \sqrt{\frac{\gamma_- - 1}{\gamma_- - \gamma_+}} \, , 1-\frac{\gamma_+}{\gamma_-} \right)\\[2mm]
        &- \nu_\theta\, \mathcal{G}_\theta(\theta_s) \,,
    \end{aligned}
\end{gather}
where we used $a>0$.
Inserting $\alpha=0$, this simplifies to
\begin{align} \label{eq:tau_is_zero_2}
\tau_0 = \frac{1}{a\sqrt{\gamma_-}} \,K\left( \frac{\gamma_- - 1}{\gamma_-} \right) -  \nu_\theta\, \mathcal{G}_\theta(\theta_s) \,.
\end{align}
Based on the periodicity of $\dn(u,m)$, we can deduce that the sign of the right-hand side of equation \eqref{eq:theta_sol_uncorrected} needs to be flipped for
\begin{align}
\frac{1+4n}{a\sqrt{\gamma_-}} \, &K\left( \frac{\gamma_- - 1}{\gamma_-} \right) - \nu_\theta\, \mathcal{G}_\theta(\theta_s) \nonumber \\
&< \, \tau \, < \frac{3+4n}{a\sqrt{\gamma_-}} \, K\left( \frac{\gamma_- - 1}{\gamma_-} \right) - \nu_\theta\, \mathcal{G}_\theta(\theta_s) \,,
\end{align}
with $n\in\R$.
Such a modified trajectory is presented in Fig.~\ref{fig:example_traj_theta_combined} (solid blue line).
The hemisphere does not change since $h=\mathrm{sign}(\cos\theta)=\mathrm{sign}(\cos (-\theta))$.
The modified analytic solution coincides with the numerical solution (cf. Sec.~\ref{sec:num-sol}), confirming the necessity of this correction.

%%%%%%%%%%%%%%%%%%%%%%%%%%%%%%%%%%%%%%%%%%%%%%%%%%%%%%%%%%%%%%%%%%%%%%%%%%%%%%%%%%%%%%%%%%%%%
\subsubsection{\texorpdfstring{$\psi$}{psi}-integral} \label{subsec:psi-sol}
To get a solution for the $\psi$-integral, it is useful to write \eqref{eq:int_eq_of_mot_EF_Mino_psi} in the form
\begin{align} \label{eq:psi_sol_1}
\psi_o = \psi_s + I_\varphi - I_{\varphi, reg}
+ \lambda \, G_\varphi
\end{align}
where we defined
\begin{align}
I_\varphi &:= \nu_r  \int_{r_s}^{r_o} \frac{ a(2mr-a\lambda)}{\Delta \sqrt{R(r)}}\, \diff r \, , \label{eq:I_phi}\\
I_{\varphi, reg} &:= \nu_r \int_{r_s}^{r_o} \frac{a}{\Delta}\, \diff r \, ,\label{eq:I_reg_phi}\\
G_\varphi &:= \nu_\theta \int_{\theta_s}^{\theta_o} \frac{ \diff \theta}{\sin^2\theta \sqrt{\Theta(\theta)}} \, . \label{eq:G_phi}
\end{align}
We can follow \cite{gralla_null_2020} and rewrite \eqref{eq:I_phi} as
\begin{align} \label{eq:I_phi_expanded}
I_\varphi = \frac{2 ma}{r_+-r_-} \Biggl[ \biggl( r_+ - \frac{a\lambda}{2m} \biggr) I_+ - \biggl( r_- - \frac{a\lambda}{2m} \biggr) I_- \Biggr] \, ,
\end{align}
where $I_\pm$ is defined in \eqref{eq:I_pm}.
The explicit calculation is given in Appendix~\ref{appsec:partial_fraction_decomposition}.
Consequently, solving for the azimuthal angle $\psi_o$ requires the antiderivatives of \eqref{eq:I_reg_phi}, \eqref{eq:G_phi}, and \eqref{eq:I_pm}.

\begin{figure}[t!]\centering
\includegraphics[width=1.0\columnwidth, keepaspectratio]{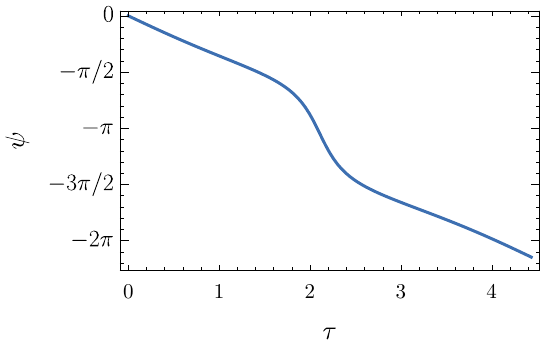}
\caption[Example for azimuthal angle trajectory]{\label{fig:example_traj_psi} An example azimuthal angle trajectory for a null geodesic with $(\alpha=-0.15, \beta=0.1)$, $r_s=+\infty$, $r_o=-\infty$ and $\theta_o=\pi/4$. The black hole's rotation parameter is set to $a/m=0.99$.}
\end{figure}

The antiderivative of \eqref{eq:I_reg_phi} is readily checked to be
\begin{align} \label{eq:mathcal_I_reg_phi}
\mathcal{I}_{\varphi,reg} = \nu_r \, \frac{a \, (\mathrm{ln}(r-r_+)-\mathrm{ln}(r-r_-))}{r_+-r_-} \, .
\end{align}
Furthermore, as shown in \cite{gralla_null_2020}, \eqref{eq:G_phi} equates to
\begin{widetext}
\begin{align} \label{eq:G_phi_sol}
G_\varphi =& \frac{1}{(1-\gamma_-) \sqrt{\gamma_-\, a^2}} \, \Pi \left( \frac{\gamma_+ - \gamma_-}{1-\gamma_-}, \Upsilon_\tau ,\, 1-\frac{\gamma_+}{\gamma_-} \right) - \nu_\theta \, \mathcal{G}_\varphi (\theta_s) \,,
\end{align}
with
\begin{align}
\Upsilon_\tau &= \am \left( \sqrt{\gamma_-\, a^2} \, \left(\tau + \nu_\theta\, \mathcal{G}_\theta (\theta_s)\right) ,\, 1-\frac{\gamma_+}{\gamma_-} \right) \, , \label{eq:Upsilon_tau}\\
\mathcal{G}_\varphi (\theta) &= -\frac{h}{(1-\gamma_-)\sqrt{\gamma_- \,a^2}} \, \Pi \left( \frac{\gamma_+-\gamma_-}{1-\gamma_-} ,\, \arcsin \sqrt{\frac{\cos^2\theta-\gamma_-}{\gamma_+-\gamma_-}} ,\, 1-\frac{\gamma_+}{\gamma_-} \right) \, , \label{eq:mathcal_G_phi}
\end{align}
\end{widetext}
and where $\gamma_\pm$ and $\mathcal{G}_\theta (\theta)$ are given by \eqref{eq:u_pm_alpha_beta} and \eqref{eq:mathcal_G_theta}, respectively.
However, we could not confirm the antiderivative of \eqref{eq:I_pm} presented in \cite{gralla_null_2020} when cross-checking it with numerical integration.
Thus, we calculated $\mathcal{I}_\pm$ anew with the help of \textit{Mathematica}~\cite{wolfram_research_inc_mathematica_nodate}.
The result can be found in \eqref{eq:I_pm_antider}-\eqref{eq:mathcal_I_2} in Appendix~\ref{appsec:antiderivatives}.
The azimuthal angle of the source $\psi_s$ can be computed using \eqref{eq:psi_sol_1} together with \eqref{eq:mathcal_I_reg_phi}, \eqref{eq:G_phi_sol} and \eqref{eq:I_pm_antider}.
An example azimuthal trajectory can be found in Fig.~\ref{fig:example_traj_psi}.

%%%%%%%%%%%%%%%%%%%%%%%%%%%%%%%%%%%%%%%%%%%%%%%%%%%%%%%%%%%%%%%%%%%%%%%%%%%%%%%%%%%%%%%%%%%%%
\subsubsection{\texorpdfstring{$u$}{u}-integral} \label{subsec:u-sol}
To solve \eqref{eq:int_eq_of_mot_EF_Mino_u}, we rewrite it as
\begin{align} \label{eq:u_sol_1}
u_o = u_s + I_t - I_{t, reg} + a^2 G_t \, ,
\end{align}
where we defined
\begin{align}
I_t &:= \nu_r \int_{r_s}^{r_o} \frac{r^2 \Delta + 2mr (r^2+a^2-a\lambda)}{\Delta \sqrt{R(r)}} \,\diff r \, , \label{eq:I_t_1}\\
I_{t, reg} &:= \nu_r \int_{r_s}^{r_o} \frac{r^2+a^2}{\Delta} \,\diff r \, , \label{eq:I_reg_t}\\
G_t &:= \nu_\theta \int_{\theta_s}^{\theta_o} \frac{ \cos^2\theta}{\sqrt{\Theta(\theta)}} \,\diff \theta \, . \label{eq:G_t}
\end{align}
As before, we can rewrite $I_t$ as in \cite{gralla_null_2020} to get
\begin{align} \label{eq:I_t_2}
I_t =& \frac{(2m)^2}{r_+-r_-} \left[ r_+ \left( r_+ - \frac{a\lambda}{2m} \right) I_+ - r_- \left( r_- - \frac{a\lambda}{2m} \right) I_- \right] \nonumber\\[2mm]
& + (2m)^2\, I_0 + 2m\, I_1 + I_2 \, ,
\end{align}
where $I_\pm$ and $I_\ell$ (with $\ell=0,1,2$) are defined in \eqref{eq:I_pm} and \eqref{eq:I_l}, respectively.
The explicit calculation can be found in Appendix~\ref{appsec:partial_fraction_decomposition}.

\begin{figure}[t!]\centering
\includegraphics[width=1.0\columnwidth, keepaspectratio]{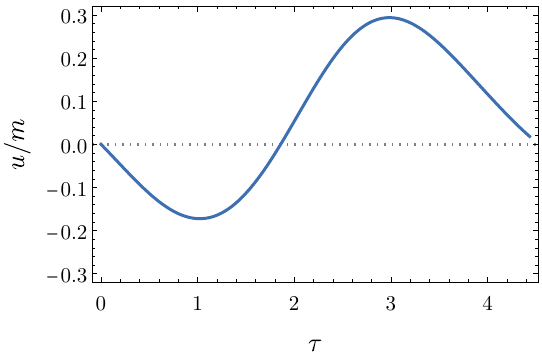}
\caption[Example for $u$-trajectory]{ An example $u$-trajectory for a geodesic with $(\alpha=-0.15,\, \beta=0.1)$, $r_s=+\infty$, $r_o=-\infty$ and $\theta_o=\pi/4$. The black hole's rotational parameter is set to $a/m=0.99$.}
\label{fig:example_traj_u}
\end{figure}

To get a solution for $u_o$ requires solutions to \eqref{eq:I_pm}, \eqref{eq:I_l}, \eqref{eq:I_reg_t}, as well as \eqref{eq:G_t}.
The antiderivative of \eqref{eq:I_pm} is given by \eqref{eq:I_pm_antider}.
Furthermore, it is straightforward to check that the antiderivative of \eqref{eq:I_reg_t} is
\begin{align}
\mathcal{I}_{t,reg} (r) =& \nu_r \Big(r + \frac{a^2 + r_-^2}{r_- - r_+} \,\mathrm{ln}(r-r_-) \nonumber\\
& \hspace{10mm}- \frac{a^2 + r_+^2}{r_- - r_+} \,\mathrm{ln}(r-r_+)\Big) \, . \label{eq:mathcal_I_reg_t}
\end{align}
Next, we can use the solution of \eqref{eq:G_t} as presented in \cite{gralla_null_2020}, for which they found
\begin{align} \label{eq:G_t_sol}
G_t = \sqrt{\frac{\gamma_-}{a^2}} \,E\left( \Upsilon_\tau ,\, 1-\frac{\gamma_+}{\gamma_-} \right) - \nu_\theta \,\mathcal{G}_t (\theta_s) \, ,
\end{align}
where $\Upsilon_\tau$ is defined as in \eqref{eq:Upsilon_tau}, and
\begin{align}
\mathcal{G}_t (\theta) = -h \,\sqrt{\frac{\gamma_-}{a^2}} \,E\left( \arcsin \sqrt{\frac{\cos^2\theta - \gamma_-}{\gamma_+-\gamma_-}} ,\, 1-\frac{\gamma_+}{\gamma_-} \right) \,.
\end{align}

Lastly, we need the antiderivative of \eqref{eq:I_l} for $\ell=0,1,2$.
For $\ell = 0$, the solution is \eqref{eq:antider_I0}.
For $\ell = 1,2$, respective formulas are presented in \cite{gralla_null_2020}.
However, we could not confirm these results, and consequently, we calculated $\mathcal{I}_1$ and $\mathcal{I}_2$ anew.
The results are presented in \eqref{eq:mathcal_I_1} and \eqref{eq:mathcal_I_2} in Appendix~\ref{appsec:antiderivatives}.
An example trajectory using \eqref{eq:u_sol_1} can be found in Fig.~\ref{fig:example_traj_u}.

%%%%%%%%%%%%%%%%%%%%%%%%%%%%%%%%%%%%%%%%%%%%%%%%%%%%%%%%%%%%%%%%%%%%%%%%%%%%%%%%%%%%%%%%%%%%%
\subsection{Numerical solution} \label{sec:num-sol}
For numerically solving the system \eqref{eq:diff_eq_of_mot_EF_Mino_r}-\eqref{eq:diff_eq_of_mot_EF_Mino_u}, we consider the second-order differential equations for $r$ and $\theta$:
\begin{align}
\frac{\diff^2 r}{\diff \tau^2} = \frac{1}{2} \frac{\diff R(r)}{\diff r} \, , \qquad
\frac{\diff^2 \theta}{\diff \tau^2} = \frac{1}{2} \frac{\diff \Theta(\theta)}{\diff \theta}\, ,
\end{align}
This form will avoid possible problems when encountering turning points in the radial or polar directions.
Furthermore, we remove possible infinities from the radial equation by making a coordinate change to $\Tilde{r} = \arctan(r)$, thereby identifying $r=\pm\infty$ with $\Tilde{r}=\pm\pi/2$.
This leads to the following system of differential equations:
\begin{align}
\frac{\diff^2 \Tilde{r}}{\diff \tau^2} &= \cos{\Tilde{r}}\,\left[ \left((a-\lambda)^2 + \eta\right) \cos{3\Tilde{r}}\right.
\\
&\left.\, + \sin{\Tilde{r}} \left(1 + a^2 \eta + \left((1+\eta)(1-a^2) + \lambda^2\right)\cos{2\Tilde{r}}\right)\right] \,,
\\
\frac{\diff^2 \theta}{\diff \tau^2} &= \lambda^2 \frac{\cos{\theta}}{\sin^3{\theta}} - a^2 \cos{\theta} \sin{\theta} \,,
\\
\frac{\diff \psi}{\diff \tau} &= \frac{a}{\Delta(\tan{\Tilde{r}})} \left( 2 \tan{\Tilde{r}} - a \lambda - \sqrt{R(\tan{\Tilde{r}})} \right)  + \frac{\lambda}{\sin^2{\theta}} \,,
\\
\frac{\diff u}{\diff \tau} &= \frac{\tan{\Tilde{r}} + a^2}{\Delta(\tan{\Tilde{r}})} \left( \tan^2{\Tilde{r}} +a^2 -a \lambda -\sqrt{R(\tan{\Tilde{r}})} \right) 
\\ 
&\, + a \lambda - a^2 +a^2 \cos^2{\theta} \,.
\end{align}

The radial potential $R(r)$ is defined in \eqref{eq:rad_pot_null}.
Moreover, we made the functional dependence in $\Delta(r) = r^2 - 2mr + a^2$ explicit and set $m=1$ everywhere.
The constants of motion $\lambda$ and $\eta$ are to be viewed as functions of the impact parameters $\alpha$, $\beta$, and $\theta_o$ by virtue of \eqref{eq:impact_param_alpha_beta_inverse}.

To solve this system numerically, we impose six initial conditions (two for each second-order equation and one for each first-order equation).
The four initial values are the positional coordinates $(u_o, \Tilde{r}_o, \theta_o, \psi_o)$ of the observer.
We set the Mino time $\tau=0$ at the observer, thus we fix $\Tilde{r}(0)=-\pi/2$, $\theta(0)=\theta_o$, $\psi(0)=\psi_o$, $u(0)=u_o$.
For each second-order equation, we additionally need to state the rate of change of the coordinate at the observer, given by $\Tilde{r}'(0)=\nu_r$ and $\theta'(0)=\beta$.
We perform the numerical integration using \textit{Mathematica}'s implementation of the explicit Bogacki-Shampine adaptive Runge-Kutta method \cite{wolfram_research_inc_mathematica_nodate}.
Because we start the integration at the observer at $\Tilde{r}_o=-\pi/2$, we will stop it once we reach the source at $\Tilde{r}_s=\pi/2$.
A comparison of the numerical and analytical solutions is presented in Appendix~\ref{appsec:ana_vs_num}.

%%%%%%%%%%%%%%%%%%%%%%%%%%%%%%%%%%%%%%%%%%%%%%%%%%%%%%%%%%%%%%%%%%%%%%%%%%%%%%%%%%%%%
\section{Visualizations} \label{sec:visualizations}

Using the presented solutions to the geodesic equations, we can inspect various vortical null geodesics in more detail.
In the following, we consider a black hole with a rotation parameter $a/m = 0.99$ and integrate from the position of the observer $(u_o = 0, \, r_o = -\infty, \, \theta_o = \pi/4, \, \psi_o =0)$ to a test source located at $r_s=+\infty$.
These values are chosen so that the inner throat remains significantly large while maximizing the asymmetry due to the rotation of the central object.
Lowering the value $a/m$ only decreases the size of the observer's field of view and renders the effect of the rotation on the null geodesics less prominent while not altering the overall qualitative behavior of null geodesics considered in this section.
Note that by inverting the parameter $\tau$ along the geodesic, we obtain a trajectory that originates at the source and terminates at the observer.

In this section, for presentation, we use Kerr-Schild coordinates, which have an underlying flat metric, and we treat the coordinates $(x,y,z)$ as pseudo-Cartesian coordinates.
Furthermore, going from one asymptotically flat region to the other (i.e., from $r>0$ to $r<0$ or vice versa) requires traversing the disk enclosed by the ring singularity \cite{chrusciel_structure_2020}.
In Kerr-Schild coordinates, the vortical null geodesics we consider can cross over from positive to negative values of $z$ only by traversing the disk in the $x$-$y$-plane, thereby changing the sign of $r$.
Thus, in the following plots, the region with positive/negative $z$ corresponds to the region with positive/negative radius $r$.
These two regions correspond to the northern hemisphere in their respective asymptotically flat ends of the spacetime (defined by the polar angle $\theta\in[0,\pi/2]$).
Consequently, the horizons are only drawn at positive $z$ while the causality-violating region is only present at negative $z$.

%%%%%%%%%%%%%%%%%%%%%%%%%%%%%%%%%%%%%%%%%%%%%%%%%%%%%%%%%%%%%%%%%%%%%%%%%%%%%%%%%%%%%%%%%
\subsection{Individual trajectories in Kerr-Schild coordinates}

%%%%%%%%%%%%%%%%%%%%%%%%%%%%%%%%%%%%%%%%%%%%%%%%%%%%%%%%%%%%%%%%%%%%%%%%%%%%%%%%%%%%%%
\subsubsection{Principal null congruence}

\begin{figure*}[p]
\centering
\includegraphics[width=1.0\textwidth, keepaspectratio]{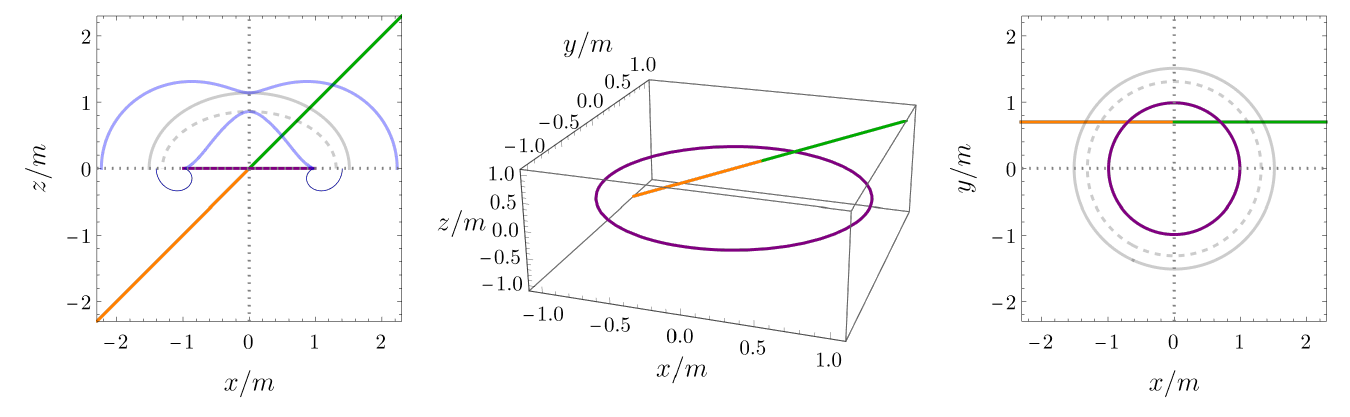}
    \caption[Trajectory of a vortical null geodesic (principal null congruence) with $\theta$ at $(\alpha=+a \sin\theta_o,\, \beta=0)$]{
    \label{fig:3d_trajs_const_theta}
    The trajectory of the principal null geodesic in Kerr-Schild coordinates for $a/m=0.99$ and $\theta_o=\pi/4$.
  It crosses the $y$-axis at $a \sin \theta_o$.
  The green part of the geodesic is at positive $r$ ($z>0$), while the orange part is at negative $r$ ($z<0$).
  The ring singularity is depicted in purple.
  The left plot is the projection onto the $x$-$z$-plane in which the upper (lower) half-plane corresponds to $r>0$ ($r<0$).
  Hence, the horizons (grey) and the ergoregion (light blue) are only drawn for $z>0$, and the causality-violating region (dark blue) is only present for $z<0$.
  The right plot is the projection onto the $x$-$y$-axis.}
\end{figure*}

The pair $(\alpha= a \sin \theta_o,\, \beta=0)$ corresponds to the principal null geodesic of the Kerr spacetime.
Along such a geodesic, the coordinates $\theta$, $\psi$, and $u$ are constant (see Sec.~\ref{sec:theta_principal_null_congruence} and Appendix~\ref{appsec:principal_null_congruence_psi_and_u}).
With the transformation law \eqref{eq:trafo_EF_to_KS}, it can be verified that this corresponds to a straight line in Kerr-Schild coordinates.
The principal null geodesic in these coordinates is plotted in Fig.~\ref{fig:3d_trajs_const_theta}.

\begin{figure*}[p]
\centering
\includegraphics[width=1.0\textwidth, keepaspectratio]{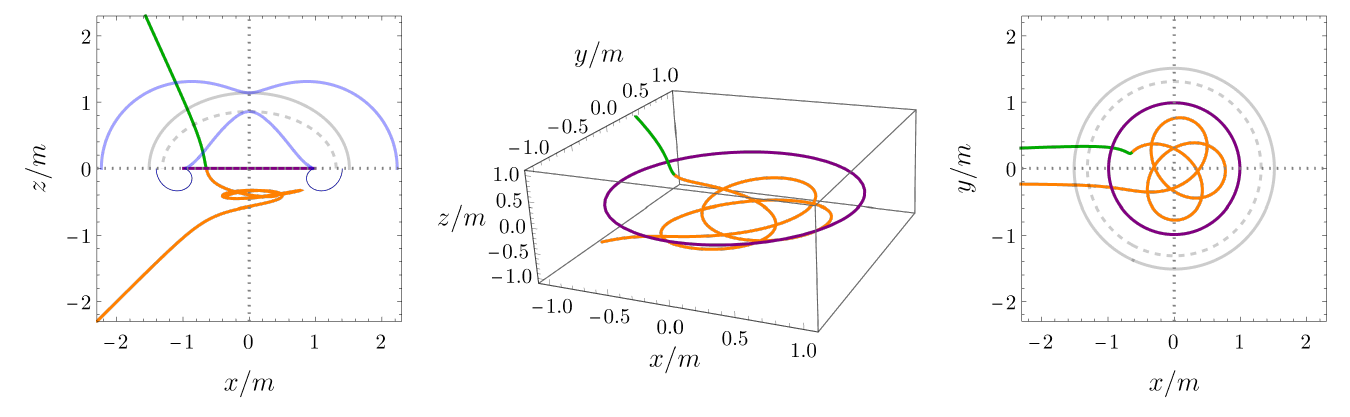}
    \caption[Trajectory of a vortical null geodesic close to the left edge of the inner throat]{
    \label{fig:3d_trajs_close_to_bnd}
    The trajectory of a vortical null geodesic close to the left inner throat boundary with $(\alpha=-0.2315214984,\, \beta=0)$ for $a/m=0.99$ and $\theta_o=\pi/4$.
  It does not intersect the causality-violating region.}
\end{figure*}

\begin{figure*}[p]
\centering
\includegraphics[width=1.0\textwidth, keepaspectratio]{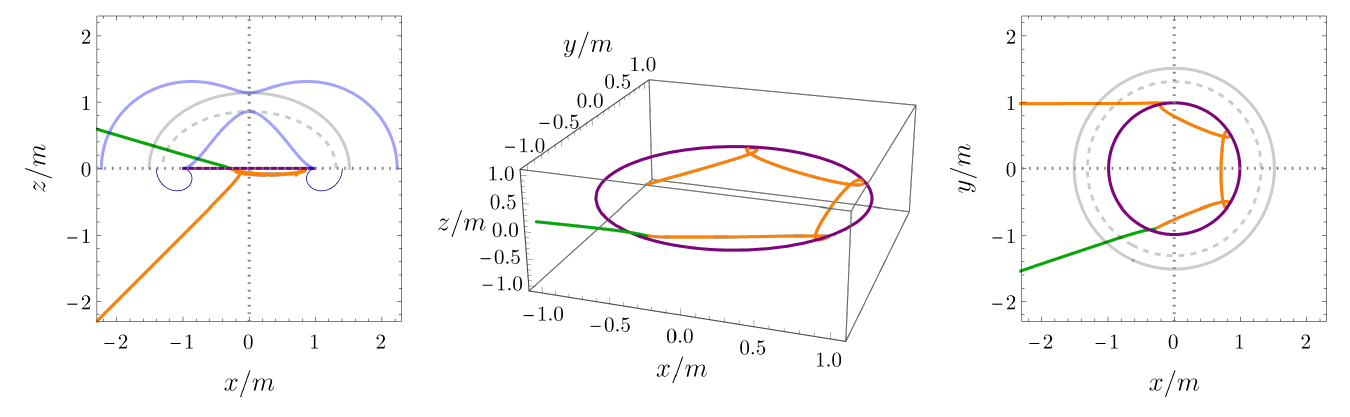}
    \caption[Trajectory of a vortical null geodesic close to the right vertex of the inner throat]{
    \label{fig:3d_trajs_close_to_bnd2}
    The trajectory of a vortical null geodesic close to the right inner throat boundary with $(\alpha=0.9783392085,\, \beta=0)$ for $a/m=0.99$ and $\theta_o=\pi/4$. It intersects the causality-violating region, see Fig.~\ref{fig:causviol_close_to_bnd2}.}
\end{figure*}

%%%%%%%%%%%%%%%%%%%%%%%%%%%%%%%%%%%%%%%%%%%%%%%%%%%%%%%%%%%%%%%%%%%%%%%%%%%%%%%%%%%%%%
\subsubsection{Geodesics close to the inner throat boundary}

Here, we address null geodesics close to the inner throat boundary (approximately $m\cdot 10^{-10}$ away).
A geodesic close to the left vertex is shown in Fig.~\ref{fig:3d_trajs_close_to_bnd}.
Its main feature is that it remains close to the ring singularity in the region with negative $r$ for a significant portion of its journey.
This period features several revolutions around the axis of symmetry and multiple polar oscillations between $\theta_+$ and $\theta_-$.
However, it never crosses the causality-violating region.

For the right vertex of the inner throat, a corresponding trajectory in Kerr-Schild coordinates is presented in Fig.\,\ref{fig:3d_trajs_close_to_bnd2}.
The $\psi$-coordinate of this geodesic is not monotonically increasing or decreasing.
In the exterior of a Kerr black hole, the \textit{order} of null geodesics can be defined as the number of times it crosses the equatorial plane from the emitter to the observer \cite{cunningham_optical_1972}.
Equivalently, one can consider the number of turning points in the polar angle along the trajectory \cite{james_gravitational_2015}.
Only the latter of these two definitions is meaningful for vortical null geodesics, which can never cross the equatorial plane.
With this, the geodesic in Fig.~\ref{fig:3d_trajs_close_to_bnd} is of order four, while the one in Fig.~\ref{fig:3d_trajs_close_to_bnd2} is of order three.

\begin{figure}[t]
\centering
\includegraphics[width=1.0\columnwidth, keepaspectratio]{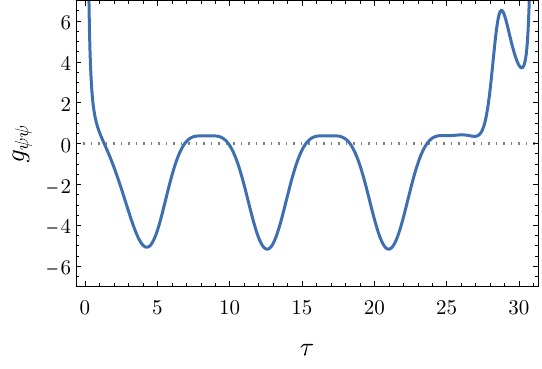}
\caption[Causality violation close to right vertex.]{\label{fig:causviol_close_to_bnd2} The metric component $g_{\psi\psi}$ along the geodesic shown in Fig.~\ref{fig:3d_trajs_close_to_bnd2}.
  As $g_{\psi\psi}$ becomes negative, the geodesic travels through the causality-violating region.}
\end{figure}

Furthermore, the geodesic close to the right vertex comes close to the ring singularity (up to $10^{-4}m$) but can never hit it \cite{chrusciel_structure_2020}.
However, it traverses the causality-violating region.
To see this, we inspect the value of the metric component $g_{\psi\psi}$ along the geodesic.
Fig.~\ref{fig:causviol_close_to_bnd2} shows the resulting plot.
Since there are regions with $g_{\psi\psi}<0$, this geodesic passes through the causality-violating region multiple times along its path.

%%%%%%%%%%%%%%%%%%%%%%%%%%%%%%%%%%%%%%%%%%%%%%%%%%%%%%%%%%%%%%%%%%%%%%%%%%%%%%%%%%%%%%%%%
\subsection{Observing the sky through the disk}

%%%%%%%%%%%%%%%%%%%%%%%%%%%%%%%%%%%%%%%%%%%%%%%%%%%%%%%%%%%%%%%%%%%%%%%%%%%%%%%%%%%%%
\subsubsection{Field of view} \label{sec:FOV}

\begin{figure*}[!ht]
  \centering
  \captionsetup[subfigure]{margin={1.3cm,0cm}}%
  \subfloat[][\label{fig:FOV_InnerThroat} ]{\includegraphics[width=0.3\textwidth]{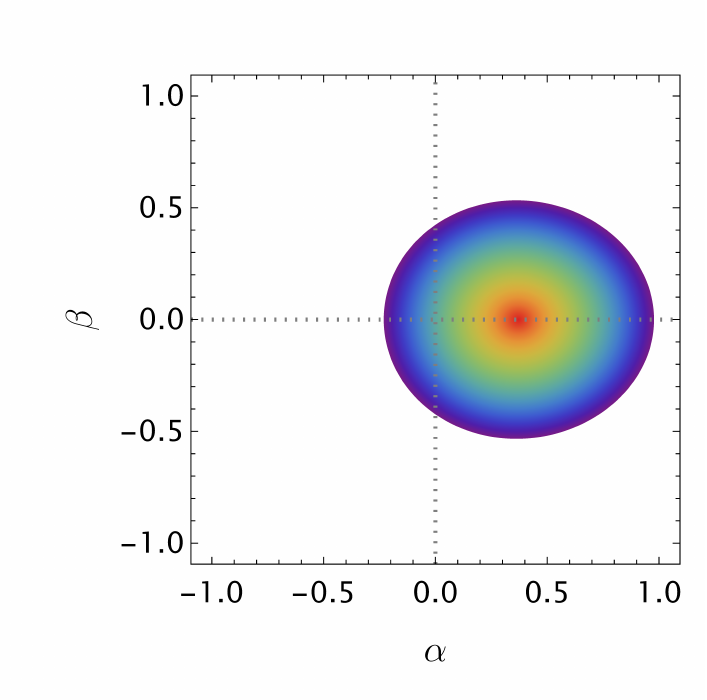}}
  \qquad
  \captionsetup[subfigure]{margin={1cm,0cm}}%
  \subfloat[][\label{fig:FOV_ThetaPsiPlot} ]{\includegraphics[width=0.3\textwidth]{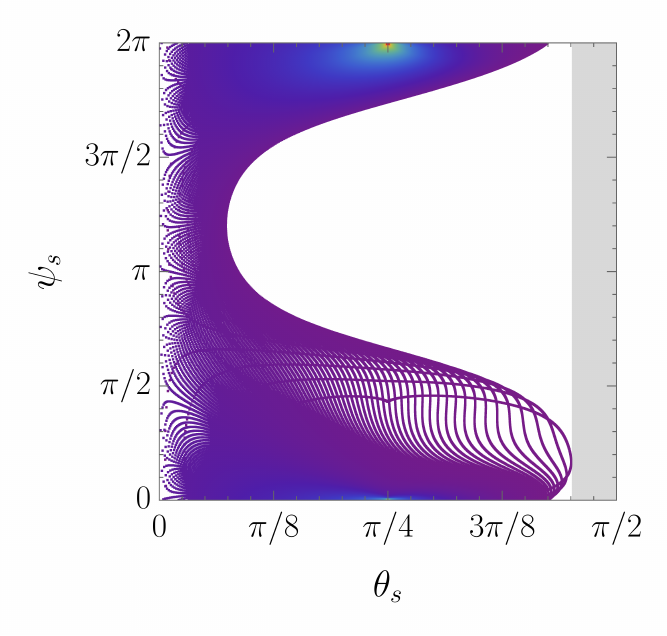}}
  \qquad
  \captionsetup[subfigure]{margin={1cm,0cm}}%
  \subfloat[][\label{fig:FOV_PolarPlot} ]{\includegraphics[width=0.3\textwidth]{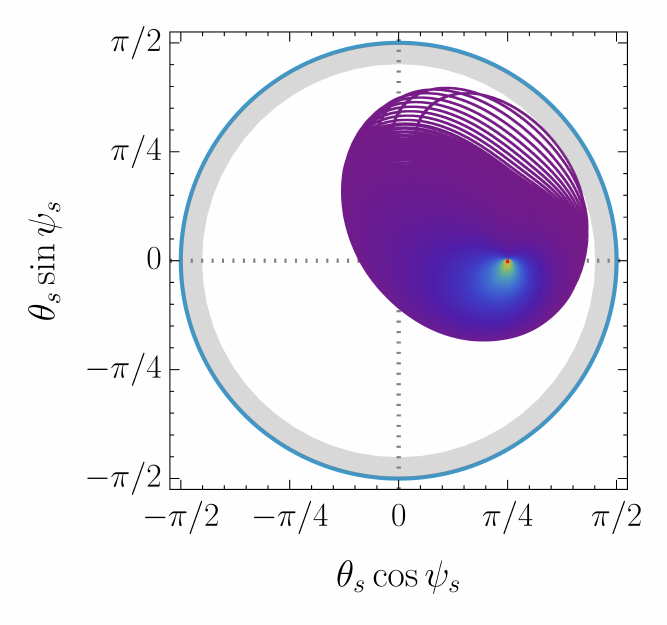}}
  \caption[The field of view for an observer at $(u_o = 0 ,\, r_o=-\infty,\,\theta_o=\pi/4,\, \psi_o=0)$]{\label{fig:FOV_combined} The field of view for an observer at $(u_o = 0 ,\,r_o=-\infty,\,\theta_o=\pi/4,\, \psi_o=0)$. \textbf{(a):} The inner throat as seen by the observer. The minimal distance to the inner throat boundary is $3\times10^{-4}$.
  \textbf{(b):} The calculated pairs of source angles $(\theta_s,\, \psi_s)$ for geodesics starting at $r_s=+\infty$. The grey area is a forbidden region from where no null geodesics can arrive at the observer.
  \textbf{(c):} Polar plot of the calculated source angles $(\theta_s,\, \psi_s)$. The distance to the origin is the polar angle $\theta_s$ while the angle going around the origin in the anticlockwise direction is $\psi_s$. The blue circle with radius $\pi/2$ corresponds to the equatorial plane; the forbidden region is now an annulus with thickness $\pi/2-\mathrm{max}(\theta_s)$.}
\end{figure*}

With our solutions to the equations of motion for vortical null geodesics, we can furthermore answer the question of what portion of the sky at $r_s=+\infty$ is visible for an observer at  $(u_o = 0,\, r_o=-\infty,\, \theta_o=\pi/4,\, \psi_o=0)$.
To this end, we discretize the inner throat, resulting in almost $1.5\cdot 10^6$ pairs of impact parameters $(\alpha,\, \beta)$ with the closest distance to the inner throat boundary of $3m\cdot 10^{-4}$.
We calculate $\theta_s$ and $\psi_s$ for each of these parameters using our analytic solutions.
To better track these pairs, we color-code them based on their distance to the center of the inner throat -- warm colors are in the middle, while cool colors correspond to the region close to the inner throat boundary (see Fig.~\ref{fig:FOV_InnerThroat}).

Fig.~\ref{fig:FOV_ThetaPsiPlot} shows the calculated starting angles in a $\theta_s$-$\psi_s$-plot, where we have already considered the azimuthal coordinate modulo $2\pi$ and where each point is colored according to its position within the inner throat.

Large portions of the sky at $r_s=+\infty$ are not visible to the observer.
This is further underpinned by the fact that Fig.~\ref{fig:FOV_ThetaPsiPlot} corresponds only to half the sky at $r_s=+\infty$ (only the northern hemisphere).
Moreover, there is a strip close to the equatorial plane
from where no photons reach the observer.
This maximal $\theta_s$ point lies at the inner throat's left edge.
The shaded grey area in Fig.~\ref{fig:FOV_ThetaPsiPlot} represents this forbidden region (cf. Sec.~\ref{sec:polar_angle_close_to_equator_or_axis_of_symmetry}).

To better visualize what portion of the sky is covered by these source positions, we make a polar plot where the distance to the origin is the polar coordinate $\theta_s$ while the angle around the origin (increasing in the anticlockwise direction) is the azimuthal angle $\psi_s$.
This corresponds to the transformation $(\theta_s,\, \psi_s)\rightarrow(\theta_s \cos(\psi_s),\, \theta_s \sin(\psi_s))$.
The resulting Fig.~\ref{fig:FOV_PolarPlot} can thus be seen as the projection of the northern hemisphere at $r_s=+\infty$ onto the two-dimensional equatorial plane.
The shaded area again indicates the forbidden regions from which no null geodesics can reach the observer.
From this polar plot, we deduce that the visible region of the sky at $r_s= +\infty$ is even smaller than suggested by Fig.\,\ref{fig:FOV_ThetaPsiPlot}.
This stems from the fact that the field of view covers the region around the axis of symmetry at $\theta_s=0$.
Furthermore, it becomes apparent that most of the inner throat (the middle part in Fig.\,\ref{fig:FOV_InnerThroat}) corresponds to a small portion of the sky at $r_s = +\infty$.
This visible region is almost centered around $(\theta_s=\pi/4,\, \psi_s=0)$ -- the same values as at the observer -- and slightly shifted in the negative azimuthal direction.
Only by approaching the boundary region of the inner throat (the purple area in Fig.~\ref{fig:FOV_InnerThroat}) does the field of view increase in size, thus ``fanning out'', and shift in the positive $\psi$-direction.

A further feature of the resulting curves in the polar plot is that the curves with a constant distance to the inner throat boundary develop caustic-like structures before they fold in on themselves at specific distances.
Due to the overlap of curves, this is not visible in Figs. \ref{fig:FOV_ThetaPsiPlot} and \ref{fig:FOV_PolarPlot}.
However, it can be observed in a short animation available at \cite{noauthor_httpswwwquantagonatmasters-thesis_nodate_www}.
We decrease the distance to the inner throat from frame to frame, thus approaching the boundary until the analytic and numerical solutions cease to coincide up to an acceptable margin (cf. Appendix~\ref{appsec:ana_vs_num}).
It is visible that the field of view shifts in the anticlockwise direction in the polar plot as the corresponding curve in the inner throat approaches the boundary, thus covering large portions of the sky at $r_s = +\infty$.
The fact that the curve in the polar plot folds in on itself results in the field of view covering the same part of the sky more than once.
Consequently, the observer perceives multiple images of the same radiating object at $r_s=+\infty$ (as in the exterior case of a Kerr black hole \cite{james_gravitational_2015}).

%%%%%%%%%%%%%%%%%%%%%%%%%%%%%%%%%%%%%%%%%%%%%%%%%%%%%%%%%%%%%%%%%%%%%%%%%%%%%%%%%%%%%%%%%
\subsubsection{Observing the sky} \label{sec:obs_sky}

A similar question concerns the appearance of a fixed test light source at $r_s = +\infty$ (e.g., the celestial sphere around the black hole) as seen by an observer at $(u_o = 0, \, r_o = -\infty, \, \theta_o = \pi/4, \, \psi_o =0)$.
For an initial study of the distortion effects (rotation, reflection, stretching, or compression of the image), we consider two sample light sources illuminating the sky at $r_s = +\infty$.
We argue in Appendix~\ref{appsec:gravitational_reshift} that we do not need to consider any gravitational redshift when the photons are emitted at $r_s = +\infty$ and observed at $r_o = -\infty$.

\begin{figure*}[!ht]
  \centering
  \captionsetup[subfigure]{margin={1.8cm,0cm}}%
  \subfloat[][\label{fig:ThetaDir_InnerThroat} ]{\includegraphics[width=0.45\linewidth]{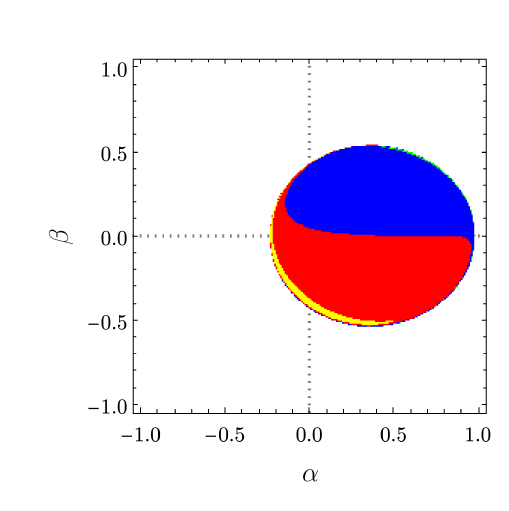}}
  \hspace{4em}
  \subfloat[][\label{fig:ThetaDir_PolarPlot} ]{\includegraphics[width=0.45\linewidth]{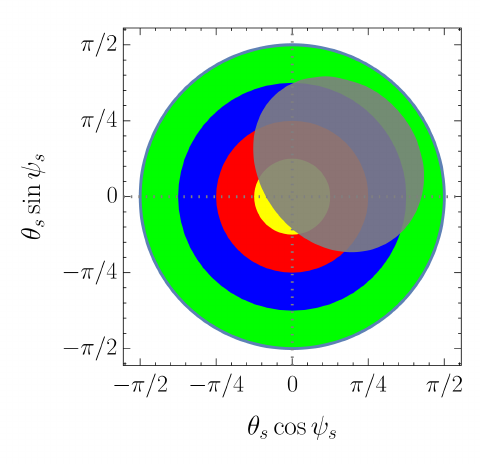}}
  \qquad
  \subfloat[][\label{fig:PsiDir_InnerThroat} ]{\includegraphics[width=0.45\linewidth]{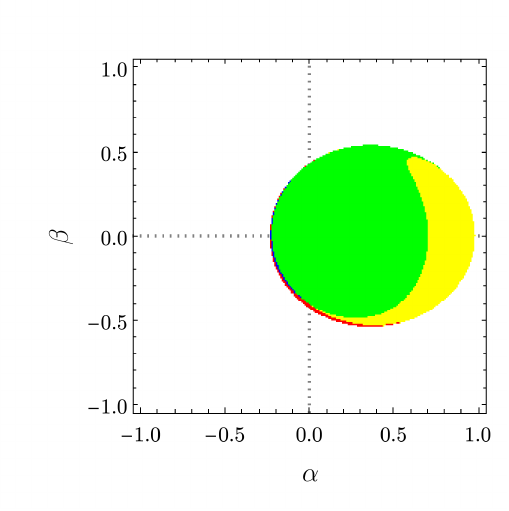}}
  \hspace{4em}
  \subfloat[][\label{fig:PsiDir_PolarPlot} ]{\includegraphics[width=0.45\linewidth]{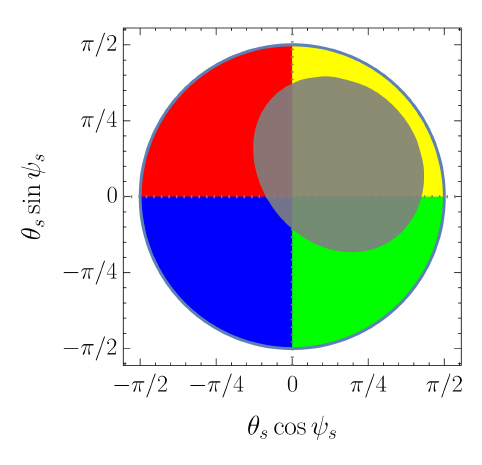}}
  \caption[Distortion of the sky at $r_s = +\infty$]{\label{fig:Distortion_combined}
  Distortion of the sky at $r_s = +\infty$ with alternating color in the polar direction (a and b) and the azimuthal direction (c and d).
  \textbf{(a):} Illustration of how an observer at $r_o=-\infty$ perceives the sky at $r_s=+\infty$ in this scenario.
  \textbf{(b):} Polar plot of the sample light source, including the calculated angles $(\theta_s,\, \psi_s)$ for the field of view in grey.
  \textbf{(c):} Illustration of how an observer at $r_o=-\infty$ perceives the sky at $r_s=+\infty$ in this scenario.
  \textbf{(d):} Polar plot of the sample light source, including the calculated angles $(\theta_s,\, \psi_s)$ for the field of view in grey.}
\end{figure*}

We first consider a light source at $r_s=+\infty$ emitting light in four basic colors depending on the polar angle $\theta_s$.
In the range $0\leq\theta_s<\pi/8$ the sky emits yellow light, for $\pi/8\leq\theta_s<\pi/4$ it emits red light, for $\pi/4\leq\theta_s<3\pi/8$ it emits blue light, and for $3\pi/8\leq\theta_s<\pi/2$ it emits green light.
The color pattern is illustrated in Fig.~\ref{fig:ThetaDir_PolarPlot}, where we overlaid the source with the calculated angles $(\theta_s,\, \psi_s)$.
Fig.~\ref{fig:ThetaDir_InnerThroat} illustrates the appearance of the inner throat for the observer at $r_o=-\infty$ in this scenario.
Each pair of impact parameters $(\alpha,\, \beta)$ is colored according to the respective position of the source angles $(\theta_s,\, \psi_s)$ on the sky at $r_s=+\infty$.
The most prominent colors, as seen by the observer, are red and blue, taking up 96.5\% of the inner throat surface area (red $\approx50\%$, blue $\approx46.5\%$); only around 3\% are yellow, and 0.5\% are green.
These shares do not change significantly when considering geodesics closer to the inner throat boundary, as their relative share within the inner throat is negligible.
As the transition from red to blue at $r_s = +\infty$ appears at $\theta_{s}=\pi/4$, the change in color within the inner throat is precisely at the point of the principal null geodesic at $(\alpha=+a \sin\theta_o,\, \beta=0)$.

Concerning distortion effects, we see that the differently colored sections are not bordered by straight lines, unlike how an observer in a flat spacetime somewhere at positive radii would perceive them.
Furthermore, the colors are skewed in one direction due to the asymmetry arising from the rotation of the black hole.
In contrast, the ordering of the colors is flipped, so, e.g., red is now ``below'' blue (compared to what an observer in flat spacetime at $r_o>0$ would see).

To study the distortion in the azimuthal direction, we consider a sample light source at $r_s=+\infty$, emitting light in different colors depending on the azimuthal angle.
The splits are as follows: for $0\leq \psi_s<\pi/2$ the sky emits yellow light, for $\pi/2\leq\psi_s<\pi$ it is red, for $\pi\leq\psi_s<3\pi/2$ it is blue, and for $3\pi/2\leq\psi_s<2\pi$ it is green (see Fig.~\ref{fig:PsiDir_PolarPlot}).
An observer at $r_o >0$ in a flat spacetime ($m=0$) would perceive this light source as follows: by turning in the positive azimuthal direction (anti-clockwise), they would see the colors of the sky changing in the order yellow - red - blue - green.
Fig.~\ref{fig:PsiDir_InnerThroat} visualizes the inner throat as seen by the observer at $r_o = -\infty$ in this scenario.
The relative size of the green and yellow regions combined is 98.4\% (green $\approx73.7\%$, yellow $\approx24.7\%$); red ($\approx1.2\%$) and blue ($\approx0.4\%$) are barely visible.
Because the center of the field of view is slightly shifted in the negative azimuthal direction (cf. Fig.~\ref{fig:FOV_ThetaPsiPlot}), the center of the inner throat is green.
Red and blue areas are restricted to the boundary region within the inner throat.

Finally, let us mention the distortion effects.
As before, the borders separating the different sections are no straight lines (as seen by an observer in a flat spacetime at positive radii) but instead curved.
Furthermore, the ordering of the colors is also flipped, e.g., green is now ``left'' of yellow.
We conclude that for an observer at $r_o=-\infty$, the source appears distorted and flipped in the polar and azimuthal direction.

%%%%%%%%%%%%%%%%%%%%%%%%%%%%%%%%%%%%%%%%%%%%%%%%%%%%%%%%%%%%%%%%%%%%%%%%%%%%%%%%%%%%%
\section{Summary}

The present paper aimed to obtain insights into the behavior of null geodesics with a global fly-by orbit, i.e., crossing from one asymptotically flat region with $r\gg 0$ to the other with $r\ll 0$.
From the emission point to the observer, they cross both horizons and the disk enclosed by the ring singularity.
Such geodesics are called vortical null geodesics, and they have a negative Carter constant.

To solve this task, we took advantage of rewriting the constants of motion in terms of impact parameters.
For an observer substantially far away from the black hole in the negative-$r$-region, they parametrize a portion of their celestial sphere, and as such, can be used to describe what the observer sees if they look at the center of symmetry of the Kerr spacetime.
In this parameter space, there is a closed region (the inner throat) for which the radial potential has no real-valued roots.
Hence, null geodesics with such impact parameters have no radial turning point.
The only photons that reach the observer at $r_o=-\infty$ from a source at $r_{s}=+\infty$ lie inside the inner throat.

An analysis of the angular potential showed that there are, at most, two distinct geodesics with a constant polar angle inside the inner throat.
One is always visible and corresponds to the principal null direction on which the Eddington-Finkelstein-like coordinates are based.
As such, it also has constant $\psi$ and $u$.
The other one is only inside the inner throat for specific values of the rotation $a/m$ and the observer's polar angle $\theta_o$.
Moreover, vortical null geodesics can never cross the equatorial plane $\theta=\pi/2$.
Consequently, an observer at $r_o=-\infty$ will only see at most half of the celestial sphere at $r_s=+\infty$.
In addition, there is a further forbidden region around the equatorial plane from which such an observer receives no photons, see Fig.~\ref{fig:theta_pm_right_edge} and \cite{noauthor_httpswwwquantagonatmasters-thesis_nodate}.

To visualize individual trajectories and simulate the observer's view, we solved the geodesic equations of motion for vortical null geodesics inside the inner throat in Eddington-Finkelstein-like coordinates analytically and numerically, with the numerical integration providing an independent validation of the analytical solution.
This way, we could correct some formulas in \cite{gralla_null_2020}.

Using our solution to the geodesic equations, we computed several distinct vortical null geodesics and visualized them in Kerr-Schild coordinates. In particular, this revealed the existence of geodesics near the inner throat boundary that may traverse the causality-violating region on their path from $r_s = +\infty$ to $r_o = -\infty$.
We then determined the starting positions corresponding to a finely discretized inner throat, allowing us to construct the field of view for an observer located at $r_o = -\infty$, $\theta_o = \pi/4$, and $\psi_o = 0$. As in the exterior Kerr case, infinitely many copies of the light source appear in the resulting image, clustering near the boundary of the inner throat.
Finally, when considering two representative light sources, we find that such an observer would perceive the sky at $r_s=+\infty$ as both distorted and flipped in the azimuthal and polar directions.

It would be interesting to extend our analysis to the case where the source is an astrophysical object orbiting the Kerr black hole in the exterior region at a finite distance $r_s>0$.
Such an object must not be a thin accretion disk: An observer at $r_o=-\infty$ could not see it because of the forbidden polar region around the equatorial plane.
However, any luminous object away from the equatorial plane would serve the purpose.
To create realistic simulations for these non-stationary sources, one must account for the gravitational redshift and the frequency shift.

Lastly, our solutions are also valid in a white hole scenario, with an observer at $r_o=+\infty$ and the source at $r_s=-\infty$.
Thus, they can also be used to simulate the appearance of a Kerr white hole with light sources located at $r_s<0$, offering a means to explore potential observational signatures, should such objects exist.

%%%%%%%%%%%%%%%%%%%%%%%%%%%%%%%%%%%%%%%%%%%%%%%%%%%%%%%%%%%%%%%%%%%%%%%%%%%%%%%%%%%%%
\begin{acknowledgments}
We thank Piotr Chru\'{s}ciel for valuable discussions and comments. This work is based on the Master's thesis of the second author \cite{noauthor_httpswwwquantagonatmasters-thesis_nodate}.
This research was funded in whole or in part by the Austrian Science Fund (FWF) [\href{http://doi.org/10.55776/P36455}{10.55776/P36455}, \href{http://doi.org/10.55776/Y963}{10.55776/Y963}, \href{http://doi.org/10.55776/COE1}{10.55776/COE1}].
\end{acknowledgments}

\appendix

%%%%%%%%%%%%%%%%%%%%%%%%%%%%%%%%%%%%%%%%%%%%%%%%%%%%%%%%%%%%%%%%%%%%%%%%%%%%%%%%%%%%%
\section{\texorpdfstring{$\psi$}{psi} and \texorpdfstring{$u$}{u} for Principal Null Geodesics} \label{appsec:principal_null_congruence_psi_and_u}

As a geodesic with $(\alpha= +a \sin\theta_o ,\, \beta=0)$ corresponds to the principal null congruence on which the transformation to Eddington-Finkelstein-like coordinates is based on (cf. \cite{poisson_relativists_2004}), we claim that along this geodesic also $\psi$ and $u$ remain constant, which we prove below.
Let us begin by evaluating the constants of motion \eqref{eq:impact_param_alpha_beta_inverse} at these impact parameters, leading to
\begin{align}\label{eq:impact_param_alpha_beta_inverse_principal_null_geod}
\lambda = a \sin^2\theta_o \, , \quad
\eta = - a^2 \cos^4\theta_o \, ,
\end{align}
with which the radial potential \eqref{eq:rad_pot_null} becomes
\begin{align}
R(r) =(r^2+a^2\cos^2\theta_o)^2 \, . \label{eq:18.03.2021_03}
\end{align}
For vortical geodesics we have $r^2+a^2\cos^2\theta_o>0$ which simplifies \eqref{eq:diff_eq_of_mot_EF_Mino_r} to
\begin{equation}
\frac{\diff r}{\diff \tau} = \nu_r (r^2+a^2\cos^2\theta_o) \, .
\end{equation}
Furthermore, we get for \eqref{eq:diff_eq_of_mot_EF_Mino_psi}
\begin{align}
\frac{\diff \psi}{\diff \tau}  &= \frac{a}{\Delta} \left(2 m r -a \lambda - \sqrt{R(r)} \right) +\frac{\lambda}{\sin^2\theta_o} \nonumber\\
&= \frac{a}{\Delta} \left(2 m r -a^2 \sin^2\theta_o - r^2 - a^2\cos^2\theta_o \right) +\frac{a \sin^2\theta_o}{\sin^2\theta_o} \nonumber\\
&= \frac{a}{\Delta} \big( 2mr - a^2 (\sin^2\theta_o + \cos^2\theta_o) \nonumber\\
& \hspace{20mm} - r^2 + (r^2-2mr+a^2) \big) \nonumber\\
&= 0 \, , \label{eq:27.04.2021_01}
\end{align}
where we used that $\theta=\theta_o$ stays constant throughout the trajectory.
Similarly, the calculation for \eqref{eq:diff_eq_of_mot_EF_Mino_u} results in
\begin{align}
\frac{\diff u}{\diff \tau} &= \frac{r^2+a^2}{\Delta} \left(r^2+a^2-a \lambda - \sqrt{R(r)} \right) + a(\lambda - a \sin^2\theta) \nonumber \\
&= \frac{r^2+a^2}{\Delta} \left( r^2 + a^2 - a^2\sin^2\theta_o - r^2 - a^2 \cos^2\theta_o \right) \nonumber\\
& \qquad+ a (a\sin^2\theta_o - a\sin^2\theta_o) \nonumber\\
&= 0 \, . \label{eq:27.04.2021_02}
\end{align}
The results \eqref{eq:27.04.2021_01} and \eqref{eq:27.04.2021_02} show that besides the polar angle $\theta$, also the azimuthal angle $\psi$, as well as the coordinate time $u$ stay constant along a geodesic with impact parameters $(\alpha= +a \sin\theta_o ,\, \beta=0)$.

%%%%%%%%%%%%%%%%%%%%%%%%%%%%%%%%%%%%%%%%%%%%%%%%%%%%%%%%%%%%%%%%%%%%%%%%%%%%%%%%%%%%%
\section{Definitions for the antiderivatives} \label{appsec:roots_and_definitions}

To calculate the antiderivatives in Sec.~\ref{sec:ana-sol}, we define additional quantities similar to \cite{gralla_null_2020}.

In the case of no radial turning point, all roots of the radial potential $R(r)$ are complex with $r_1=r_2^*$ and $r_3=r_4^*$ (where the star denotes complex conjugation).
These roots are given by \cite{gralla_null_2020}
\begin{align}
r_1 = -z - \sqrt{-\frac{\mathcal{X}}{2}-z^2 + \frac{\mathcal{Y}}{4z}} \,, \\
r_2 = -z + \sqrt{-\frac{\mathcal{X}}{2}-z^2 + \frac{\mathcal{Y}}{4z}} \,, \\
r_3 = -z - \sqrt{-\frac{\mathcal{X}}{2}-z^2 - \frac{\mathcal{Y}}{4z}} \,, \\
r_4 = -z + \sqrt{-\frac{\mathcal{X}}{2}-z^2 - \frac{\mathcal{Y}}{4z}} \,,
\end{align}
with
\begin{align}
\mathcal{X} &= a^2 - \eta - \lambda^2 \,,\\
\mathcal{Y} &= 2m \left( \eta + (\lambda - a)^2 \right) \,,\\
\mathcal{Z} &= -a^2 \eta \,, \label{eq:17.03.2021_01}\\
z &= \sqrt{\frac{\xi_0}{2}} \,,\\
\xi_0 &= \omega_+ + \omega_- - \frac{\mathcal{X}}{3} \,,\\
\omega_\pm &= \sqrt[3]{-\frac{\mathcal{Q}}{2} \pm \sqrt{\left( \frac{\mathcal{P}}{3} \right)^3 + \left( \frac{\mathcal{Q}}{2} \right)^2}} \,,\\
\mathcal{P} &= -\frac{\mathcal{X}^2}{12} - \mathcal{Z} \,,\\
\mathcal{Q} &= -\frac{\mathcal{X}}{3} \left[ \left( \frac{\mathcal{X}}{6} \right)^2 - \mathcal{Z} \right] - \frac{\mathcal{Y}^2}{8} \,.
\end{align}
Furthermore, some additional formulas needed in Sec.~\ref{sec:ana-sol} are
\begin{align}
r_3 &= b_1 - i a_1 \, ,\\
r_1 &= b_2 - i a_2 \, ,\\
C &= \sqrt{(r_3-r_1)(r_4-r_2)} \, , \label{eq:17.03.2021_02}\\
D &= \sqrt{(r_3-r_2)(r_4-r_1)} \, ,\\
g_0 &= \sqrt{\frac{4a_2^2 - (C-D)^2}{(C+D)^2-4a_2^2}} \, ,\\
k &= \frac{D^2}{C^2} \, ,\\
k_4 &= \frac{4 C D}{(C+D)^2} \, ,\\
x_4(r) &= \frac{r+b_1}{a_2} \, ,\\
h &= \mathrm{sign}(\cos\theta_s) \, ,
\end{align}
where $a_1, b_1, a_2, b_2$ are uniquely specified by the restrictions $r_1 = r_2^\ast$ and $r_3 = r_4^\ast$ on the roots.
For the calculation of the radial solution, we additionally need
\begin{equation}
X_4(\tau) = \frac{C+D}{2} \, (\nu_r \tau + \mathcal{I}_0(r_s)) \, ,
\end{equation}
where $\mathcal{I}_0(r_s)$ denotes the antiderivative of \eqref{eq:I_l} for $\ell=0$ evaluated at the source's radial position $r_s$, and $\tau$ is the Mino time.

%%%%%%%%%%%%%%%%%%%%%%%%%%%%%%%%%%%%%%%%%%%%%%%%%%%%%%%%%%%%%%%%%%%%%%%%%%%%%%%%%%%%%%%%%%%%%
\section{Elliptic Integrals}\label{subsec:elliptic_integrals}
For completeness, this section provides a basic introduction to elliptic functions following \cite{abramowitz_handbook_1972} and \cite{gradshtein_table_2000}.
An integral of the form $\int R(x,y) \, \diff x$ is called an elliptic integral if $R(x,y)$ is a rational function and $y=\sqrt{P(x)}$ where $P(x)$ is a polynomial either cubic or quartic in $x$.
Generally, these integrals cannot be solved using elementary functions.
All elliptic integrals can be reduced to integrals of elementary functions and three canonical elliptic integrals:
\begin{align}
F(\varphi,m) &:= \int_0^\varphi \frac{\diff \theta}{\sqrt{1-m \sin^2\theta}} \,, \label{eq:ell_int_1_incomplete}\\
E(\varphi,m) &:= \int_0^\varphi \sqrt{1-m \sin^2\theta}\, \diff \theta \,,\label{eq:ell_int_2_incomplete}\\
\Pi(n,\varphi,m) &:= \int_0^\varphi \frac{\diff \theta}{(1-n \sin^2\theta) \sqrt{1-m \sin^2\theta}} \, .\label{eq:ell_int_3_incomplete}
\end{align}
These are known as the elliptic integrals of the first, second, and third kind, respectively.
The canonical elliptic integrals are all odd functions in $\varphi$.
In the general case $0<\varphi<\pi/2$, the integrals \eqref{eq:ell_int_1_incomplete}-\eqref{eq:ell_int_3_incomplete} are called incomplete canonical elliptic integrals.
For $\varphi=\pi/2$, they are known as complete canonical elliptic integrals and denoted by $K(m):= F\left(\pi/2,m\right)$, $E(m):= E\left(\pi/2,m\right)$, and $\Pi(n,m):= \Pi\left(n,\pi/2,m\right)$.

Based on the elliptic integrals, one can define inverse functions as follows.
Let $u=F(\varphi,m)$ be the elliptic integral of the first kind. We define
\begin{equation}
\am(u,m):=\varphi
\end{equation}
to be the inverse of $F(\varphi,m)$.
The function $\am(u,m)$ is called Jacobi amplitude, and by definition it satisfies $F\left(\am(u,m),m\right)=u$.
From this, three further functions follow:
\begin{align}
\sn(u,m) =& \sin(\am(u,m)) =\sin(\varphi) \, ,\\
\cn(u,m) =& \cos(\am(u,m)) =\cos(\varphi) \, ,\\
\dn(u,m) =& \sqrt{1-m \sin^2(\am(u,m))} \, ,
\end{align}
which are called sine-amplitude, cosine-amplitude, and delta-amplitude, respectively.
From the definition it follows immediately that ${\sn^2(u,m)+\cn^2(u,m)=1}$ and ${\dn^2(u,m)+m\,\sn^2(u,m)=1}$.
The functions $\sn(u,m)$ and $\cn(u,m)$ are periodic in $u$ with period $4K(m)$, whereas $\dn(u,m)$ is $2K(m)$-periodic.
In the following we also need $\mathrm{sc}(u,m)= \sn(u,m) /\cn(u,m)$.

%%%%%%%%%%%%%%%%%%%%%%%%%%%%%%%%%%%%%%%%%%%%%%%%%%%%%%%%%%%%%%%%%%%%%%%%%%%%%%%%%%%%%%%%
\section{Rewriting \texorpdfstring{$I_\varphi$}{Iphi} and \texorpdfstring{$I_t$}{It}} \label{appsec:partial_fraction_decomposition}
To simplify \eqref{eq:I_phi} we use $\Delta=(r-r_+)(r-r_-)$ to write
\begin{align}
I_\varphi &= \nu_r \int_{r_s}^{r_o} \frac{\diff r}{\sqrt{R(r)}} \, \frac{a(2mr - a\lambda)}{(r-r_+)(r-r_-)} \nonumber\\
&= \nu_r \int_{r_s}^{r_o} \frac{\diff r}{\sqrt{R(r)}} \, f_1(r) \label{eq:app17.03.2021_01}
\end{align}
and use partial fraction decomposition to rewrite $f_1(r)$.
Utilizing the equations
\begin{gather}
    \begin{aligned} \label{eq:partial_frac_decomp_1}
        \frac{1}{(r-r_+)(r-r_-)} &= \frac{1}{(r-r_+)(r_+-r_-)}\\
        & \hspace{4mm}-\frac{1}{(r-r_-)(r_+-r_-)} \, ,\\[4pt]
        \frac{r}{(r-r_+)(r-r_-)} &= \frac{r_+}{(r-r_+)(r_+-r_-)}\\
        & \hspace{4mm}-\frac{r_-}{(r-r_-)(r_+-r_-)}
    \end{aligned}
\end{gather}
we obtain
\begin{align*}
f_1(r)= \frac{2ma}{r_+-r_-} \left( \frac{ r_+ - a\lambda/2m}{r-r_+} - \frac{r_- - a \lambda/2m}{r-r_-} \right) \,.
\end{align*}
Inserting this and \eqref{eq:I_pm} into \eqref{eq:app17.03.2021_01}, we find
\begin{align*}
I_\varphi = \frac{2ma}{r_+-r_-} \Biggl[ \biggl( r_+ - \frac{a\lambda}{2m} \biggr) I_+ - \biggl( r_- - \frac{a\lambda}{2m} \biggr) I_- \Biggr] \, .
\end{align*}

Simplifying \eqref{eq:I_t_1} is very similar.
First, let us write $I_t$ as
\begin{align}
I_t &= \nu_r \int_{r_s}^{r_o} \frac{\diff r}{\sqrt{R(r)}} \, \left( r^2 + \frac{2mr (r^2+a^2-a\lambda)}{(r-r_+)(r-r_-)} \right) \nonumber\\
&= \nu_r \int_{r_s}^{r_o} \frac{\diff r}{\sqrt{R(r)}} \, f_2(r) \,. \label{eq:app17.03.2021_04}
\end{align}
Using the partial fraction decompositions \eqref{eq:partial_frac_decomp_1} and
\begin{align} \label{eq:partial_frac_decomp_2}
\frac{r^3}{(r-r_+)(r-r_-)} = & r + r_+ + r_- \nonumber\\
&+ \frac{1}{r_+-r_-} \left( \frac{r_+^3}{r-r_+} - \frac{r_-^3}{r-r_-} \right) \,, 
\end{align}
we can rewrite $f_2(r)$ as
\begin{align} \label{eq:f2}
f_2(r)=& \frac{(2m)^2}{r_+-r_-} \left( \frac{r_+ (r_+ - a\lambda/2m)}{r-r_+}  - \frac{r_- (r_- - a\lambda/2m}{r-r_-} \right) \nonumber\\[2mm]
&+ r^2 + 2mr +(2m)^2  \,,
\end{align}
where we used that $r_+r_-=a^2$ and $r_+ + r_- = 2m$.
Plugging \eqref{eq:f2}, \eqref{eq:I_pm}, and \eqref{eq:I_l} into \eqref{eq:app17.03.2021_04} yields
\begin{align*}
I_t =& \frac{(2m)^2}{r_+-r_-} \left[ r_+ \left( r_+ - \frac{a\lambda}{2m} \right) I_+ - r_- \left( r_- - \frac{a\lambda}{2m} \right) I_- \right] \\[2mm]
&+ (2m)^2 I_0 + 2m I_1 + I_2 \, .
\end{align*}

%%%%%%%%%%%%%%%%%%%%%%%%%%%%%%%%%%%%%%%%%%%%%%%%%%%%%%%%%%%%%%%%%%%%%%%%%%%%%%%%%%%%%%%%
\section{Antiderivatives of \texorpdfstring{$I_\pm$}{I+-}, \texorpdfstring{$I_1$}{I1}, \texorpdfstring{$I_2$}{I2}}
\label{appsec:antiderivatives}

Here, we present the antiderivatives $I_\pm$, $I_1$, and $I_2$, defined in \eqref{eq:I_pm} and \eqref{eq:I_l}.
We calculated these anew because we could not confirm the formulas presented in \cite{gralla_null_2020} when comparing them with numerical methods.

Symbolic integration of $I_\pm$ using \textit{Mathematica} yields
\begin{widetext}
\begin{align} \label{eq:mathcal_I_pm}
\mathcal{I}^1_\pm (r) =&  \frac{\nu_r i (r_2-r_\pm)}{\sqrt{a_1 a_2} (r_1 - r_\pm)(r_3 - r_\pm)} \,  F\biggl( \arcsin \sqrt{\frac{2 i a_2 (r-r_3)}{(r-r_1)(r_2-r_3)}} ,\, \frac{D^2}{4a_1 a_2} \biggr) \nonumber\\
& + \frac{\nu_r i (r_2-r_\pm)(r_1-r_3)}{\sqrt{a_1 a_2} (r_1 - r_\pm)(r_3 - r_\pm)} \, \Pi \biggl( \frac{(r_2-r_3)(r_1-r_\pm)}{2 i a_2 (r_3-r_\pm)} ,\,  \arcsin \sqrt{\frac{2 i a_2 (r-r_3)}{(r-r_1)(r_2-r_3)}}  ,\, \frac{D^2}{4a_1 a_2} \biggr) \,.
\end{align}
This function is not continuous over the whole range of $r$.
We can correct this by introducing an integration constant of the form
\begin{align} \label{eq:mathcal_I_pm_corr_const}
c_{corr,\pm} (r)=& -2 \nu_r H[-r] \, H\left(-\mathrm{Im}\left(\arcsin \sqrt{\frac{2 i a_2 (r-r_3)}{(r-r_1)(r_2-r_3)}} \right)\right) \times \nonumber\\
&\times	\Biggl( \frac{i (r_2-r_\pm)}{\sqrt{a_1 a_2} (r_1 - r_\pm)(r_3 - r_\pm)}  K\biggl(\frac{D^2}{4a_1 a_2} \biggr)  + \frac{i (r_2-r_\pm)(r_1-r_3)}{\sqrt{a_1 a_2} (r_1 - r_\pm)(r_3 - r_\pm)} \Pi \biggl( \frac{(r_2-r_3)(r_1-r_\pm)}{2 i a_2 (r_3-r_\pm)} ,\, \frac{D^2}{4a_1 a_2} \biggr) \Biggr) \,,
\end{align}
where $H(\,\cdot\,)$ denotes the Heaviside function and $\mathrm{Im}(\,\cdot\,)$ denotes the imaginary part of the argument.
Adding this constant to \eqref{eq:mathcal_I_pm} yields a smooth solution for all values of $r$.
Thus, the antiderivative of $I_\pm$ coinciding with the numerical integration over the whole range of $r$ is
\begin{equation} \label{eq:I_pm_antider}
\mathcal{I}_\pm (r) = \mathcal{I}^1_\pm (r) + c_{corr,\pm} (r) \,.
\end{equation}
Symbolic integration of $I_1$ and $I_2$ using \textit{Mathematica} yields
\begin{align}
\mathcal{I}_1 (r) =& \nu_r \sqrt{\frac{4}{(r_2-r_4)(r_3-r_1)}} \Biggl( 2a_1 \, \Pi\biggl( \frac{r_4-r_1}{r_3-r_1} ,\, \arcsin \sqrt{\frac{(r_3-r_1)(r-r_4)}{(r_4-r_1)(r-r_3)}}  ,\, k \biggr) \nonumber\\
&\hspace{38mm} - i r_3 \, F \biggl( \arcsin \sqrt{\frac{(r_3-r_1)(r-r_4)}{(r_4-r_1)(r-r_3)}} ,\, k \biggr) \Biggr) \nonumber\\
& - 2 \nu_r H(-r) H\left( -\mathrm{Im}\left( \arcsin \sqrt{\frac{(r_3-r_1)(r-r_4)}{(r_4-r_1)(r-r_3)}} \right) \right) \mathrm{Re}\Biggl( 2a_1 \, \Pi\left( \frac{r_4-r_1}{r_3-r_1} ,\, k \right) - i r_3 \, K(k) \Biggr) \, ,  \label{eq:mathcal_I_1}\\[20pt]
\mathcal{I}_2 (r) =& \nu_r \Bigg\{ (r-r_1) \sqrt{\frac{(r-r_4)(r-r_2)}{(r-r_1)(r-r_3)}} - \sqrt{(r_3-r_1)(r_4-r_2)} \, E\left( \arcsin \sqrt{\frac{(r_3-r_1)(r-r_4)}{(r_4-r_1)(r-r_3)}} ,\, k \right) \nonumber\\
& \hspace{5mm} + \sqrt{\frac{4}{(r_4-r_2)(r_3-r_1)}} \, (b_1^2 - a_1 a_2) \, F\left( \arcsin \sqrt{\frac{(r_3-r_1)(r-r_4)}{(r_4-r_1)(r-r_3)}} ,\, k \right) \Bigg\} \nonumber\\
& - 2 \nu_r H(-r) H\left( -\mathrm{Im}\left( \arcsin \sqrt{\frac{(r_3-r_1)(r-r_4)}{(r_4-r_1)(r-r_3)}} \right) \right) \nonumber\\
& \hspace{5mm}\times  \mathrm{Re}\left( - \sqrt{(r_3-r_1)(r_4-r_2)} \, E(k) + \sqrt{\frac{4}{(r_4-r_2)(r_3-r_1)}} \, (b_1^2 - a_1 a_2) \, K(k) \right) \, , \label{eq:mathcal_I_2}
\end{align}
where the last term in each formula is the correction constant to ensure a smooth solution for all values of $r$, and $\mathrm{Re}(\,\cdot\,)$ denotes the real part of the argument.
\end{widetext}

%%%%%%%%%%%%%%%%%%%%%%%%%%%%%%%%%%%%%%%%%%%%%%%%%%%%%%%%%%%%%%%%%%%%%%%%%%%%%%%%%%%%%%%%
\section{Comparing analytics and numerics} \label{appsec:ana_vs_num}

Here, we compare the analytic solution of the geodesic equations presented in Sec.~\ref{sec:ana-sol} with the numeric solution discussed in Sec.~\ref{sec:num-sol}.
The analytic solutions are simply the difference between the antiderivatives of the integrals of motion evaluated at the start and endpoint of geodesics.
In contrast, numerical integration uses a finite step size to determine the next point in the trajectory.
This finite step size can become a problem in impact-parameter regions where slight deviations in the trajectory result in entirely different geodesic behaviors, e.g., in the vicinity of the inner throat boundary.

To determine how accurate the solutions are, we compare the results for $\theta_s$, $\psi_s$ and $u_s$ for geodesics connecting $r_s=+\infty$ to $r_o=-\infty$.
This is the maximal range in $r$; thus, the error should be most significant for such trajectories.
We consider the accuracy of the equations in the example case $a/m=0.99$ for an observer located at $\theta_o=\pi/4$, $\psi_o=0$, and $u_o=0$.
These values are chosen so that the inner throat remains sufficiently large while allowing us to examine asymmetric features arising from the black hole’s rotation.
To put the distances in the $(\alpha,\beta)$-impact-parameter-plane that we consider below into perspective, recall that in the just described setting, the inner throat is egg-shaped with a semi-major axis of around $0.6$ and semi-minor axis of about $0.5$, see Fig.~\ref{fig:inner_throats}.

\begin{figure*}[th!]\centering
  \centering
  \subfloat[][\label{fig:diff_num_ana_arctan_r} ]{\includegraphics[width=0.45\linewidth]{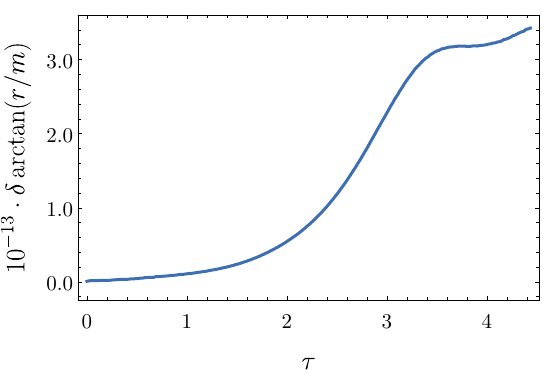}}
  \qquad
  \subfloat[][\label{fig:diff_num_ana_theta} ]{\includegraphics[width=0.45\linewidth]{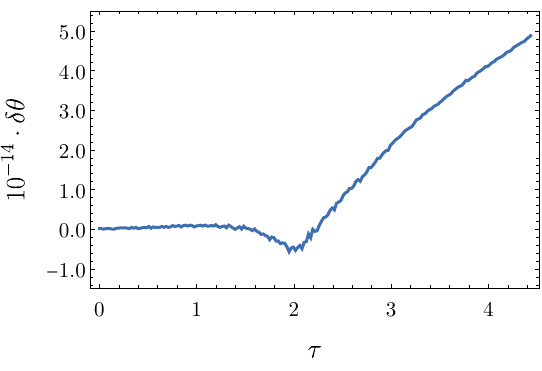}}
  \qquad
  \subfloat[][\label{fig:diff_num_ana_psi} ]{\includegraphics[width=0.45\linewidth]{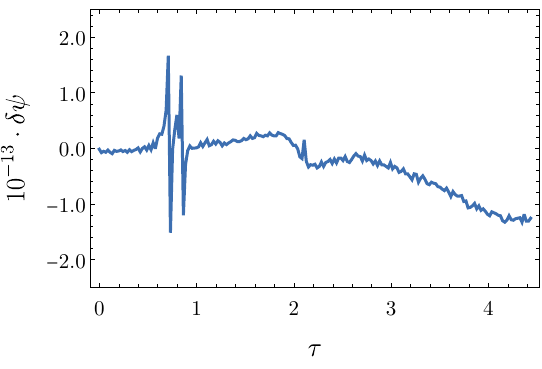}}
  \qquad
  \subfloat[][\label{fig:diff_num_ana_u} ]{\includegraphics[width=0.45\linewidth]{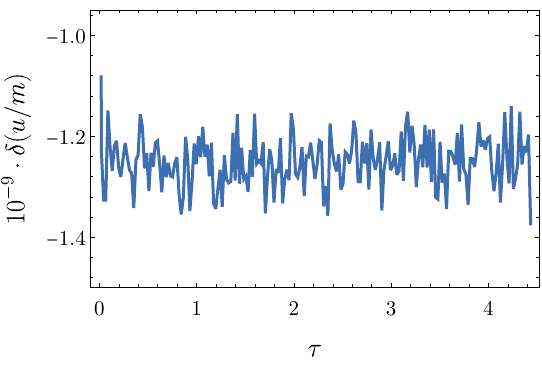}}
  \caption[Difference of the numerical and the analytic solution for an example geodesic]{\label{fig:diff_num_ana} Difference of the numerical and the analytic solution for a vortical null geodesic with $(\alpha=-0.15,\,\beta=0.1)$, $r_s=+\infty$, $r_o=-\infty$ and $\theta_o=\pi/4$. The black hole's rotation parameter is set to $a/m=0.99$. The error ranges lie within the required numerical precision.}
\end{figure*}

Far inside the inner throat (up to a distance of $m\cdot 10^{-3}$ from the inner throat boundary), the difference between the analytical and the numerical solutions for $\theta_s$ and $u_s$ is less than $10^{-9}$, while for $\tau_s$ and $\psi_s$ it is less than $10^{-8}$.
We interpret this as a complete agreement of the two solutions.
Fig.~\ref{fig:diff_num_ana} shows the difference of the numerical and analytic solution along a vortical null geodesic with impact parameters $(\alpha=-0.15,\,\beta=0.1)$, which corresponds to the example presented in the figures in Sec.~\ref{sec:ana-sol}.
As expected, the error grows with decreasing distance to the throat boundary.
At a distance of $m\cdot 10^{-8}$ from the boundary the maximal error for $\theta_s$ and $u_s$ is around $10^{-5}$, and $10^{-4}$ for $\tau_s$ and $\psi_s$.
Even closer at $m\cdot 10^{-10}$ away from the boundary, we find that the maximal disagreement in $\theta_s$, $u_s$, $\tau_s$, as well as $\psi_s$ is of order $10^{-3}$.
This leads us to conclude that the error we make in the worst case is of the order of machine precision ($\approx 10^{-16}$) divided by the distance to the throat boundary times a factor of $10^4$.

%%%%%%%%%%%%%%%%%%%%%%%%%%%%%%%%%%%%%%%%%%%%%%%%%%%%%%%%%%%%%%%%%%%%%%%%%%%%%%%%%%%%%
\section{Gravitational redshift} \label{appsec:gravitational_reshift}

To create images of what an actual physical observer might see, one task is to consider the redshift along the geodesic.
In this section, we discuss the gravitational redshift and comment on the Doppler shift due to the relative motion of the source and observer, which we want to consider.
The calculation is done using Boyer-Lindquist coordinates $(t_{BL}, r_{BL}, \theta_{BL}, \varphi_{BL})$ of the Kerr spacetime \cite{visser_kerr_2008}.
For brevity, we will denote them by $(t, r, \theta, \varphi)$ in this section.

We consider the observer to be stationary, which in Boyer-Lindquist coordinates is described by the worldline
\begin{equation}
\gamma^\mu_o (t) = (t,\, r_o,\, \theta _o,\, \varphi_o) \nonumber
\end{equation}
where $r_o$, $\theta_o$ and $\varphi_o$ are constant.
The four-velocity of this observer is given by
\begin{equation}
U_o^\mu = (v_o^t , 0 , 0 , 0) \,. \nonumber
\end{equation}
Note that this does not necessarily correspond to an observer with zero angular momentum at any radial position $r_o$, who, due to the dragging of frames by the rotation of the black hole, does not have a constant azimuthal coordinate but rather rotates with an angular velocity \cite{bardeen_timelike_1972}
\begin{equation}
\frac{\diff \varphi}{\diff t} = \frac{2 a m r_o}{(r_o^2+a^2)^2 - (r_o^2-2 m r_o +a^2) a^2 \sin^2\theta} \,. \nonumber
\end{equation}
Only in the limit $|r_o|\rightarrow\infty$ we find that a constant azimuthal coordinate corresponds to zero angular momentum of the observer.
We normalize the four-velocity by imposing 
\begin{equation}
-1 = g(U_o,U_o) = (v_o^t)^2 g_{tt} \ \Rightarrow\ v_o^t = \frac{1}{\sqrt{-g_{tt}}} \,, \nonumber
\end{equation}
and consequently, we obtain
\begin{equation}
U_o^\mu = \left(\frac{1}{\sqrt{-g_{tt}(r_o)}} ,\, 0 ,\, 0 ,\, 0\right) \,. \label{eq:16.03.2021_04}
\end{equation}
To emphasize and keep track of the fact that $g_{tt}$ needs to be evaluated at the spacetime location of the observer, we made this dependence on $r_o$ explicit.
Since $g_{tt}(r_o)=-(1-\frac{2mr_o}{\Sigma(r_o)})$, an observer at $|r_o|\gg1$ has four-velocity $U_o^\mu = (1, 0, 0, 0)$.

The light source we consider in Sec.~\ref{sec:obs_sky} is stationary and located at fixed $\theta_s$, $\psi_s$, and $r_s=+\infty$. The four-velocity $U_s$ of such a source is the same as for an observer at $r_o=-\infty$.

We calculate the gravitational redshift in the same manner as one does the calculation in the Schwarzschild case (see, e.g., \cite{chrusciel_elements_2020}).
The frequency of the emitted light, as measured by an observer with four-velocity $U_a$ ($a\in\{o,s\}$, where $o$ denotes the considered observer and $s$ denotes the light source), is given by
\begin{equation} \label{eq:16.03.2021_01}
\omega_a = -g(U_a,K) \,,
\end{equation}
where $K = \kappa \dot{\gamma}$ (where $\kappa>0$) is the frequency four-vector of a photon corresponding to a constant multiple of the tangent $\dot{\gamma} = (\dot{t},\dot{r},\dot{\theta},\dot{\varphi})$ of a null geodesic $\gamma = (t,r,\theta,\varphi)$ in Kerr spacetime (the dot denotes differentiation with respect to some affine parameter along the geodesic). Evaluating \eqref{eq:16.03.2021_01} further we find
\begin{align} \label{eq:16.03.2021_02}
\omega_a &= -g_{\mu \nu} \, U_a^\mu \, K^\nu \nonumber\\
&= -U_a^t (r_a)\, \left(g_{tt}(r_a)\, \kappa\, \dot{t} + g_{t\varphi}(r_a)\, \kappa\, \dot{\varphi}\right) \,,
\end{align}
where again we choose $a=o$ for the observer or $a=s$ for the source.
The dependence of $g_{tt}$ and $g_{t\varphi}$ on the radial coordinate is made explicit again.

To relate this frequency measured by the observer to the frequency measured at the source, i.e. relate $\omega_o$ to $\omega_s$, we make use of the fact that energy is conserved along any geodesic, which corresponds to $\partial_t$ being a Killing vector. Therefore, the quantity
\begin{equation} \label{eq:16.03.2021_03}
g(\partial_t,K)=g_{tt}\, \kappa\, \dot{t} + g_{t\varphi}\, \kappa\, \dot{\varphi}=\const\,
\end{equation}
is conserved along the null geodesic connecting the source and the observer. Using this, we find the relation of the frequency $\omega_o$ as measured by an observer with the frequency $\omega_s$ as measured at the source to be
\begin{align}
\omega_o &= -U_a^t (r_o) (g_{tt}(r_o) \kappa \dot{t} + g_{t\varphi}(r_o) \kappa \dot{\varphi}) \nonumber\\
&= -U_a^t (r_o) (g_{tt}(r_s) \kappa \dot{t} + g_{t\varphi}(r_s) \kappa \dot{\varphi}) \nonumber\\
&= \frac{U_a^t (r_o)}{U_a^t (r_s)} \, \omega_s \nonumber\\
&= \frac{\sqrt{-g_{tt}(r_s})}{\sqrt{-g_{tt}(r_o})} \, \omega_s \,,
\end{align}
where in the last line, we used \eqref{eq:16.03.2021_04}.
As already mentioned earlier, we have that $g_{tt}(r)\rightarrow -1$ as $|r|\rightarrow\infty$, and thus in the case of the observer being at $r_o=-\infty$ and the source being at $r_s=+\infty$ there will be no gravitational redshift and we find $\omega_o=\omega_s$.

\bibliography{references}% Produces the bibliography via BibTeX.

\end{document}